\documentclass[preprint,11pt]{JHEP3}

\JHEPspecialurl{http://jhep.sissa.it/JOURNAL/JHEP3.tar.gz}

\usepackage{epsfig,multicol,amsmath}

\usepackage{bm}  
\usepackage{amsmath}     

\usepackage{graphicx}
\usepackage{rotating}
\usepackage{pifont}
\usepackage{relsize}
\usepackage{mathtools}
\usepackage{cite}

\def\beq{\begin{equation}}
\def\eeq{\end{equation}}
\def\bea{\begin{eqnarray}}
\def\eea{\end{eqnarray}}

\def\eq#1{{Eq.~(\ref{#1})}}
\def\fig#1{{Fig.~\ref{#1}}}
\newcommand{\bas}{\bar{\alpha}_S}
\newcommand{\as}{\alpha_S}

\newcommand{\Lb}{\left(}
\newcommand{\Rb}{\right)}
\newcommand{\h}{\frac{1}{2}}

\newcommand{\prm}{^{\,\prime}}

\newcommand{\he}{\hspace{0.3cm}=\hspace{0.3cm}}

\parskip=\itemsep               

\newcommand{\nn}{\nonumber}
\newcommand{\D}{\partial}

\newcommand{\ga}{\gamma}

\newcommand{\Ga}{\Gamma}
\newcommand{\om}{\omega}

\newcommand{\f}{\frac}
\newcommand{\lab}{\label}

\def\pom{{I\!\!P}}

\title{BFKL Pomeron calculus: solution to equations for nucleus-nucleus scattering in the saturation domain}
\author{{\Large Carlos Contreras${}^{a}$,  Eugene Levin${}^{a,b}$ and Rodrigo Meneses${}^{c}$}
\\
${}^a$\,\, Departamento de F\'\i sica, Universidad T\'ecnica
Federico Santa Mar\'\i a \, and  \,Centro Cient\'ifico-Tecnol$\acute{o}$gico de Valpara\'\i so,  Avda. Espa\~na 1680, Casilla 110-V,  Valpara\'\i so, Chile\\
${}^b$ \, Department of Particle Physics, School of Physics and Astronomy,
Tel Aviv University, Tel Aviv, 69978, Israel\\
${}^c$\,\, Escuela de Ingenier\'\i a Civil, Facultad de Ingenier\'\i a, Universidad de Valpara\'\i so, Avda Errazuriz 1834, Valpara\'\i so, Chile\\}

\abstract{In this paper we solve the equation for nucleus-nucleus scattering in the BFKL Pomeron calculus, suggested by Braun\cite{BRA}. We find these solutions analytically at high energies as well as numerically in the entire region of energies inside the saturation region. The semi-classical approximation is used to select out the infinite set of the parasite solutions. The nucleus-nucleus cross sections at high energy are estimated and compared with the Glauber-Gribov  approach.  It turns out that the exact formula gives the estimates that are very close  to the ones based on Glauber-Gribov formula which is important  for the practical applications.  }

\dedicated{PACS: 13.85.-t, 13.85.Hd, 11.55.-m, 11.55.Bq}
\preprint{TAUP\,\, \\

\today}

\begin{document}

\section{Introduction}

The goal of this paper is to find the solution to the equations for nucleus-nucleus collision that have been derived in Ref.\cite{BRA}.
We continue the attempts, taken in Refs.\cite{BOMO,BOBR,KLM,CLM}, to study these equations and to search the general method  of solving them.

 Nucleus-nucleus scattering gives the most informative example of dense - dense  parton system interactions in which  we can see the main prediction of Color Glass Condensate/saturation approach\cite{GLR,MUQI,MV,MUCD,BK,JIMWLK}. However, in spite of the fact that we know quite well  qualitative features of nucleus-nucleus scattering (see Refs. \cite{CGCREV}) CGC/saturation approach  suffers by the absence of the evolution equation that gives us a possibility to find  the  scattering amplitude at high energy. On the other hand we know quite well the initial condition for such an evolution \cite{MV}. Fortunately, the second approach to the high energy QCD; the BFKL\cite{BFKL,RevLI} Pomeron calculus, gives the equations for nucleus-nucleus scattering at high energy. For dilute-dense parton system scattering both approaches: BFKL Pomeron calculus and Color Glass Condensate (CGC),  lead to the same nonlinear equations\cite{BRA,BAKU}.
Therefore, we can hope that the equations given by BFKL Pomeron calculus would be proven in the framework of CGC.

In the next section we give the brief review of the equations derived in Ref.\cite{BRA} and discuss their main properties. This section does not contain any  new results except section 2.3, and it is written for the  completeness of presentation. In section 2.3 we consider the asymptotic solution to the problem at large values of rapidity Y in the framework of the semiclassical approach that has been developed by us in Refs.\cite{KLM,CLM}. In the next section  we show that the number of possible solutions has to be reduce to unique solution which is discussed in section 5. In section 6 we  derive that nucleus-nucleus amplitude for the solution given in section 5.
In conclusions we summarize our results and compare our solution with the numerical solutions of Refs.\cite{BOMO,BOBR}. Unfortunately, the main equations were proposed a decade ago but we have only had five papers  devoted to a search of the solutions (see Refs.\cite{BOMO,BOBR,KLM,CLM} and this paper).

\section{The BFKL Pomeron calculus for nucleus-nucleus interaction at high energy}

\subsection{Equations for nucleus-nucleus scattering}

The most economic and elegant form  the BFKL Pomeron calculus has in terms of the functional integral \cite{BRA}
\beq 
 \label{BFKLFI}
Z[\Phi, \Phi^+]\,\,=\,\,\int \,\,D \Phi\,D\Phi^+\,e^S \,\,\,\hspace{0.5cm}\mbox{with}\hspace{0.5cm}\,S \,=\,S_0
\,+\,S_I\,+\,S_E
\eeq
where $S_0$ describes free Pomerons, $S_I$ corresponds to their mutual interaction
while $S_E$ relates to the interaction with the external sources (target and
projectile).

We will write these actions in the momentum representation\cite{BOMO,CLM} which is defined as

\bea
&&\Phi^\dag\Lb x_1,x_2,Y\prm\Rb\,\,\,\,\,\,=\,\,\,\,\,\,\Phi^\dag\Lb x_{12},b,Y\prm\Rb\,\,\,\,\,\, =\,\,\,\,\,\,x_{12}^2\int d^2k_1 e^{-ik_1\cdot x_{12}}\Phi^\dag\Lb k_1,b,Y\prm\Rb\lab{fourier}\\
&&\Phi\Lb x_1,x_2,Y\prm\Rb\,\,\,\,\,\,=\,\,\,\,\,\,\Phi\Lb x_{12},b,Y\prm\Rb\,\,\,\,\,\, =\,\,\,\,\,\,x_{12}^2\int d^2k_2 e^{ ik_2\cdot x_{12}}\Phi\Lb k_2,b,Y\prm\Rb\lab{fourier2}\eea

$S_0$ takes the form
\bea \label{S0}
&&S_0\,\,=\,\,
64\Lb 2\pi\Rb^2\int \! dY'\!\int \!d^2b\,\!\int\! d^2k\,\Phi^\dag\Lb k,b,Y\prm\Rb  \left\{ \,\Lb \f{\D}{\D l} +1\Rb^2\,\, \f{\D^2}{\D l^2}   \,\,
\Big\{\f{\D}{\D Y'} \,-\,{\cal H}\Big\} \Phi \Lb k,b,Y\prm\Rb\,\right\}
\eea
where $l \,\,=\,\,\ln k^2$ and 
\beq \label{H}
{\cal H} \Phi \Lb k,b,Y\Rb\,\,=\,\,\bas\left\{\,\int \frac{d^2 k'}{\Lb \vec{k}\,-\,\vec{k}'\Rb^2}\, \Phi \Lb k',b,Y\Rb\,\,-\,\,\h \int \frac{k^2 d^2 k'}{k'^2\,\Lb \vec{k}\,-\,\vec{k}'\Rb^2}\,\Phi \Lb k,b,Y\Rb\right\}
\eeq
where $\bas\, =\, \Lb N_c/\pi\Rb \as $ ($N_c$ is the number of colours and  the running QCD coupling $\as = 1/\Lb \beta_0 \,\ln\Lb k^2/\Lambda_{QCD}^2\Rb\Rb$ and $\beta_0 = \Lb 33 - 2\,n_f\Rb/12\pi$ with $n_f$ is the number of the fermions).

The interaction term $S_I$ can be written as follows\cite{CLM}:

 \bea\label{SI}
 &&S_I\,=\, 16 \,\f{(2\pi)^5\bas^2}{N_c}\int \!dY\prm\!\int\!4\,d^2b\int\! d^2k\, 
\left\{\,\Phi^\dag\Lb - k,b,Y\prm\Rb\Phi^\dag\Lb - k,b,Y\prm\Rb\,\Lb \f{\D}{\D l} +1\Rb^2\,\f{\D^2}{\D l^2}
\Phi\Lb k,b,Y\prm\Rb\right.\nn\\
&&\left.~~~~~~~~~~~~~~~~~~~~~~~~~~~~~~~~~~~~~~~~~~~~~ + \,\,\Phi\Lb -k, b,Y\prm\Rb \Phi\Lb - k,b,Y\prm\Rb \Lb \f{\D}{\D l} +1\Rb^2\,\f{\D^2}{\D l^2}
\Phi^\dag\Lb k,b,Y\prm\Rb\right\} 
\eea

The equations for nucleus-nucleus scattering have been derived from the averaging of the equations of motion for the  action of \eq{BFKLFI}

\beq \label{EQOM}
\Big{ \langle} \frac{\delta S}{\delta \,\Phi\Lb k, b, Y' \Rb}\Big{\rangle}\,\,=\,\,0~~~~~~~~~~~\Big{ \langle} \frac{\delta S}{\delta \,\Phi^\dagger\Lb k, b, Y'\Rb}\Big{\rangle}\,\,=\,\,0
\eeq
where averaging is understood as
\beq  \label{O}
\langle O(x,z;Y') \rangle \,\,\equiv\,\frac{\int D \Phi D \Phi^\dagger\, O(k, b,Y') \,e^{S\left[\Phi, \Phi^\dagger\right]}}{
\int D \Phi D \Phi^\dagger\, \,e^{S\left[\Phi, \Phi^\dagger\right]}|_{S_E = 0}}
\eeq

Deriving the equation of motion we assume that
\bea \label{IDTY}
  \Big{\langle} \Phi^2\Lb k, Y';b \Rb\Big{\rangle}&&\he \Lb \Big{\langle} \Phi\Lb k, Y';b \Rb\Big{\rangle}\Rb^2\lab{N2}\\
\Big{\langle} \Lb\Phi^\dag\Rb^2\Lb k, Y';b \Rb\Big{\rangle}&&\he \Lb \Big{\langle} \Phi^\dag\Lb k, Y';b \Rb\Big{\rangle}\Rb^2\nn \\
\Big{\langle} \Phi\Lb k, Y';b \Rb\,\Phi^\dag\Lb k, Y'; b\Rb\Big{\rangle} &&\he \Big{\langle} \Phi\Lb k, Y';b \Rb\Big{\rangle}\times  \Big{\langle} \Phi^\dag\Lb k, Y';b \Rb\Big{\rangle}\nn
   \eea
     These identities  are proven in the case of nucleus-nucleus scattering within accuracy of about $1/A^{1/3}$ (see Refs.\cite{BRA,KLM,BK}). We need to find the relation between fields $\Phi\Lb k, b ,Y'\Rb$ and $\Phi^\dag\Lb k, b,Y'\Rb$ and the scattering amplitude owing to the single BFKL Pomeron exchange for $S_I =0$ which we denote $N\Lb k, k_0; b,Y'\Rb$ (where $k$ and $k_0$ is the final and initial transverse momenta at rapidity $Y$  and $Y_0$, respectively).
       
Taking into account \eq{IDTY} one can see that the  variation with respect to $\Phi^\dag\Lb k, b,Y'\Rb$ leads to the following equation of motion

\bea \label{EQPHI}
&&
\delta \Lb S_0\,+\, S_I\Rb / \delta \Phi^{\dag}(k, b, Y')\he 64 (2\pi)^{2}  \,\,\Lb\frac{\partial}{\partial l}\,+\,1\Rb^2\frac{\partial^{2}}{\partial l^{2}} \, \Big(\frac{\partial}{\partial Y'} \,\,-\,\,H\Big) \Phi \Lb k,b,Y' \,\Rb\\
&&+\, 16\Lb\f{2\pi\bas^2}{N_c}\Rb\Lb 2\pi\Rb^4 \left\{  2\Phi^{\dag}\Lb -k,b,Y\prm\Rb  \Lb\frac{\partial}{\partial l}+1\Rb^2
\f{\partial^2}{\partial l^2} \Phi\Lb k,b,Y\prm\Rb +  \Lb\frac{\partial}{\partial l}+ 1\Rb^2\f{\partial^2}{\partial l^2} \Phi^2 \Lb -k,b,Y\prm\Rb\right\}\he 0\nn
\eea
 Using $S_0$ we easily see that 
   \beq \label{NN}
   N\Lb k, k_0, b; Y; Y_0\Rb\,\, =\,\, 2 \pi^2 \as \Big{\langle}\Phi(l,b,Y - Y_0 ) \Big{\rangle}   \eeq

   For understanding the relation between field $\Phi^\dag$ and $   N\Lb k, K_0, b; Y; Y_0\Rb$ where $K_0$ is the transverse momentum at rapidity $Y$ we use the equation\cite{GLR,MUSH}
   \beq \label{UN}
    N\Lb K_0, k_0, b; Y; Y_0\Rb\,\,=\,\,\,\int d^2 b' d l \,N^\dag\Lb K_0, k,\vec{b} - \vec{b}', Y -  Y'\Rb   \,N\Lb k , k_0, \vec{b}', Y' -  Y_0\Rb
    \eeq
\eq{UN}     has more general meaning than for exchange of one Pomeron (see Ref.\cite{KLM} for proof in the case of nucleus-nucleus scattering): it gives the analytical continuation of the $t$-channel unitarity at large values of energy.  For the BFKL Pomeron exchange we have
\bea \label{1P}
&& N_\pom\Lb K_0, k_0, b; Y; Y_0\Rb\,\,\,=\,\,\,N_\pom\Lb L , b,Y\Rb\,\,\, =\,\,\,\int \,\frac{d \ga}{2 \pi i}\,n_{\pom}\Lb \ga, b \Rb \,e^{\bas \chi(\ga)\,Y\,\,-\,\,( 1 - \ga)\,L}\,\,\\
&&=\,\,\int d^2 b' \int d \,l\int \,\frac{d \ga^\dag}{2 \pi i}\,\int \,\frac{d \ga}{2 \pi i}
\,n^\dag_{\pom}(\ga^\dag, \vec{b}-\vec{b}')\,e^{ \bas \chi(\ga)\,\Lb Y\,-\,Y^{\prime}\Rb\,\,-\,\,( 1 - \ga^
\dag)\,\Lb L\,-\,l\Rb}\,n_{\pom}\Lb \ga, b'\Rb e^{ \bas \chi(\ga')\,\,Y^{\prime}\,\,-\,\,(1 - \ga')\,l}\nn
\eea
where $L\,\,=\,\,\ln\Lb K^2_0/k^2_0\Rb$ and $ l\,\,=\,\,\ln\Lb k^2/k^2_0\Rb$.  $K_0$ and $k_0$ are the momenta of the dipoles in the projectile and the target, respectively. Integrating over $l$ we obtain that $\ga^\dag = \ga$. Considering $Y'=Y_0$ we obtain that
\beq \label{NDAG}
  N^\dagger\Lb  L, l, b; Y; Y_0\Rb\,\, =\,\, 2 \pi^2 \as \Big{\langle}\Phi^\dag\Lb L - l ,b,Y - Y_0 \Rb \Big{\rangle}_{S_I=0}\,\,=N\Lb b; L - l, Y - Y_0\Rb
  \eeq

Assuming that \eq{1P} and \eq{NDAG} hold in the general case but not only for the BFKL Pomeron exchange, we reduce \eq{EQPHI}
to the following equation for the amplitudes
 \bea \label{EQ}
&&0\,\,\,=\,\,\Lb\,\frac{\partial}{\partial l}\,+\,1\Rb^{2}\,\frac{\partial^{2}}{\partial l^{2}}\, \, \Lb\,\frac{\partial}{\partial Y'} \,\,-\,\,{\cal H}\Rb\, N \Lb\, l,b,Y'\,\Rb\,\,\,\\
&& \,\,+\,\bas\left\{ 2\,N\Lb\, L - l, b,Y  -  Y'\Rb\,\Lb\,\frac{\partial}{\partial l}\,+\,1\Rb^{2}\,\frac{\partial^{2}}{\partial l^{2}}\, \,\,
N\Lb\, l, b,Y'\Rb\,\,\,+\,\, \Lb\,\frac{\partial}{\partial l}\,+\, 1\Rb^{2}\,\frac{\partial^{2}}{\partial l^{2}}\, N^2 \Lb\, l, b,Y'\Rb\,\,\,\right\}
\nn
\eea
The second equation that stems from variation with respect to $\Phi\Lb l,b,Y'\Rb$ has the same form as \eq{EQ}.

In Ref.\cite{CLM} we solve \eq{EQ} in semi-classical approximation assuming that
\beq \label{SCSOL}
N\Lb l, b, Y'\Rb\,\,=\,\,e^{S\Lb l, b, Y'\Rb}\,\,=\,\,e^{\omega\Lb l, b, Y'\Rb\,Y' \,\,-\,\,\Lb 1 - \gamma\Lb l, b, Y'\Rb\Rb\,l}
\eeq
and using the method of characteristics. 
In \eq{SCSOL} we consider that  $\omega\Lb l, b, Y'\Rb\
\,=\,\partial S\Lb l, b, Y'\Rb/\partial Y'$ and $\gamma\Lb l, b, Y'\Rb\,=\,\partial S\Lb l, b, Y'\Rb/\partial l $ are smooth functions of $Y'$ and $l$ (see
Ref.\cite{CLM} for more details). We found that for any value of $z_Y$ there exists the solution at large $l$ which is very close to the solution of the linear BFKL equation. In particular this solution has a critical characteristic for $\ga = \ga_{cr}$ that can be found from the following equation \cite{GLR,MUT}

\beq \label{GACR}
\frac{\chi\Lb \ga_{cr}\Rb}{1 - \ga_{cr}}\,\,=\,\,- \frac{d \chi\Lb \ga_{cr}\Rb}{d \ga_{cr}}
\,\,\,\,\,\mbox{where}\,\,\,\,\,\,\chi\Lb \ga\Rb\,=\,2 \psi\Lb 1 \Rb \,-\,\psi\Lb \ga\Rb \,-\,\psi\Lb 1 - \ga\Rb \,\leftarrow \mbox{kernel of the BFKL equation}
\eeq
with $\psi(z) = d \ln \Ga(z)/d z$ and $\Ga(z)$ is the Euler gamma-function.

The equation for the saturation scale looks as follows
\beq \label{QS}
z \,\,\equiv\,\,\ln \Big(\frac{ Q^2_s\Lb Y'; b\Rb}{k^2}\Big)\,\,=\,\,\bas \frac{\chi\Lb \ga_{cr}\Rb}{1 - \ga_{cr}}\,Y'\,\,-\,\,l
\eeq
In the vicinity of the saturation scale but for $ z \,<\,0$ the scattering amplitude shows the geometric scaling behaviour \cite{IIM} i.e. it depends only on one variable ($z$) instead of three: $Y'$, $l$ and $b$. For the Balitsky-Kovchegov equation the geometric scaling behaviour of the scattering amplitude is the typical feature inside  the saturation region (see Ref.\cite{GS,LT}). In this paper we are going to solve \eq{EQ} treating $N\Lb z; Y'\Rb$ as a function of two variable: $z$ (see \eq{QS})  and $Y'$. The choice of the variable shows that we believe that the scattering amplitude inside the saturation region has the geometric scaling behaviour and the initial condition for this solution can be found from the solution outside of the saturation scale, namely,
\beq \label{ICZ}
N \Lb z = 0, Y' \Rb\,\,=\,\,N_0;\,\,\,\,\,\,\,\,\,\,\,\,\,\,\frac{d N\Lb z \Rb}{d z}|_{z = 0}\,\,=\,\,\Lb1 - \ga_{cr}\Rb\,N_0\,\,=\,\,1 - e^{-\phi_0}
\eeq
Recall that on the critical trajectory, the amplitude is constant and in the vicinity of the saturation scale it is proportional to $N_0\,\exp\Big( \Lb 1 - \ga_{cr}\Rb\,z\Big)$. However, introducing a dependence on $Y'$ we are going to check whether the assumption on the scaling behavior of the amplitude is correct and within what accuracy. The initial conditions at $Y'=0$ we set using the McLerran - Venugopalan formula\cite{MV} (see term $\{\dots\}$ below), namely
\beq \label{INY}
N \Lb z , Y'  =  0 \Rb\,\,=\,\,\int \frac{d^2 x_{12}}{x^2_{12}}\Big\{ 1\,\,-\,\,e^{ - Q_s\Lb Y'=0\Rb \,x^2_{12}/4}\Big\}
\,\,=\,\,\h \Gamma\Lb 0, k^2/Q^2_s\Lb Y'=0\Rb\Rb
\eeq


       \subsection{Solution inside the saturation domain: general approach}

For finding the solution inside the saturation region we will use a method proposed in Ref.\cite{LT}(see also Refs.\cite{BKL,CLM}): we introduce function $\phi\Lb z\Rb$  as follows
\beq \label{N}
N\Lb z , Y'\Rb\,\,=\,\,\h \int^z_0 \,d z' \Big( 1 \,-\,e^{- \phi\Lb z',Y'\Rb}\Big)\,\,+\,\,N_0
\eeq
and assuming that function $\D \phi\Lb z ,Y'\Rb/\D z$ is a smooth function we will find the solution to \eq{EQ}.

 The smoothness of function $\D \phi\Lb z ,Y'\Rb/\D z$ means that
 \beq \label{SMOOTH}
 \Lb \frac{\D}{\D z}\Rb^n\,N\Lb z, Y' \Rb\,\,=\,\,\h \Lb \frac{\D}{\D z}\Rb^{n - 1}\Big( 1 \,-\,e^{- \phi\Lb z, Y'\Rb}\Big)
 \,\,=\,\,\h \Lb -\,\frac{\D \phi\Lb z \Rb}{\D z}\Rb^{n - 1}e^{- \phi\Lb z, Y'\Rb}
 \eeq 
 
 Using \eq{SMOOTH} as well as the properties of the BFKL equation (see Ref.\cite{CLM} for details) we
 obtain the following equation

 \beq \label{EQD}
\tilde{\omega} \,+\, {\cal L}\Lb \gamma\Rb - {\cal F}\Lb \gamma\Rb \,e^{ - \phi\Lb z \Rb}\,\,=\,\,2 N\Lb z, Y'\Rb\,\,+\,\,2 N^\dagger\Lb z, Y'\Rb\,\,=\,\,2 N\Lb z,Y'\Rb \,\,+\,\, 2 N^\dagger\Lb z_Y  \,-\,z,  Y  - Y'\,\Rb 
\eeq

 Where
\bea
&&\gamma\,=\,\frac{\D \phi\Lb z, Y'\Rb}{ \D z}\,\,~~~~~~~~\mbox{and}~~~~~~~~~~~
\,\,\tilde{\omega}\,=\,\frac{\D \phi\Lb z, Y'\Rb}{\bas\, \D Y'}; \label{OMGA}\\
&&
\phi\Lb z= 0, Y' \Rb\,\,=\,\,(1 - \ga_{cr})\,N_0\,\,=\,\,0.63\,N_0 \,\,\,\,\mbox{and}\,\,\,\,\,\phi\Lb z, Y'=0\Rb\,\,=\,\,\phi_0\,e^z;\label{IC}\\
&&
{\cal L}\Lb \gamma\Rb \,\,=\,\,\frac{\chi\Lb \ga_{cr}\Rb}{1 - \ga_{cr}}\,\gamma\,\,+\,\,\chi\Lb \gamma\Rb \,-\,\frac{1}{\gamma} \,\,+\,\,\frac{1 + 3\gamma}{\gamma\,(1 + \gamma)};\label{LCAL}\\
&&
  {\cal F}\Lb \gamma\Rb \,\,=\,\,\frac{1\, + \,6\gamma\,+\,7 \gamma^2}{\gamma \,(1 + \gamma)^2};
\label{FCAL}
\eea  
  Introducing $\tilde{N} = 2 N$ we can rewrite \eq{EQD} in the form
   \bea \label{EQ00}
&& \, \frac{\tilde{N}''_{z Y'}\Lb z, Y' \Rb}{\Lb 1 - \tilde{N}'_z\Lb z, Y' \Rb\Rb}\,=\\
&&\,{\cal L}\Lb \frac{\tilde{N}''_{z z}\Lb z , Y'\Rb}{\Lb 1 - \tilde{N}'_z\Lb z, Y' \Rb\Rb}\Rb - {\cal F}\Lb  \frac{\tilde{N}''_{z z}\Lb z, Y' \Rb}{\Lb 1 - \tilde{N}'_z\Lb z , Y'\Rb\Rb}\Rb \,\Lb
1 - \tilde{N}'_z\Lb z, Y' \Rb\Rb\,\,-\,\,\tilde{ N}\Lb z,Y'\Rb \,\,-\,\, \tilde{ N}\Lb z_Y  \,-\,z,  Y  - Y'\,\Rb\nn 
\eea  
  
  Initial conditions  for  this equation  are given by \eq{IC}.


  \subsection{Asymptotic solution}

Demonstrating the main features and the problems that we face solving \eq{EQD} we, first, investigate the specific case: 
 $\phi\Lb z, Y' \Rb $ increases at large $z$  and it has a geometric scaling behaviour $
\phi\Lb z, Y' \Rb\,\,=\,\,\phi\Lb z \Rb$.  Using these assumptions
we  can simplify the general equation (see \eq{EQD}):
\beq \label{AS1}
{\cal L}\Lb \ga \Rb\,\,=\,\,2\,N\Lb z \Rb\,\,+\,\,2\,N\Lb z_Y - z\Rb \,\,=\,\,z_Y
\eeq
Function ${\cal L}\Lb \ga = d \phi/d z\Rb$ is shown in \fig{l} (black line). One can see that we have the following asymptotic solutions\cite{CLM}:

\FIGURE[ht]{
\begin{tabular}{c }
\epsfig{file=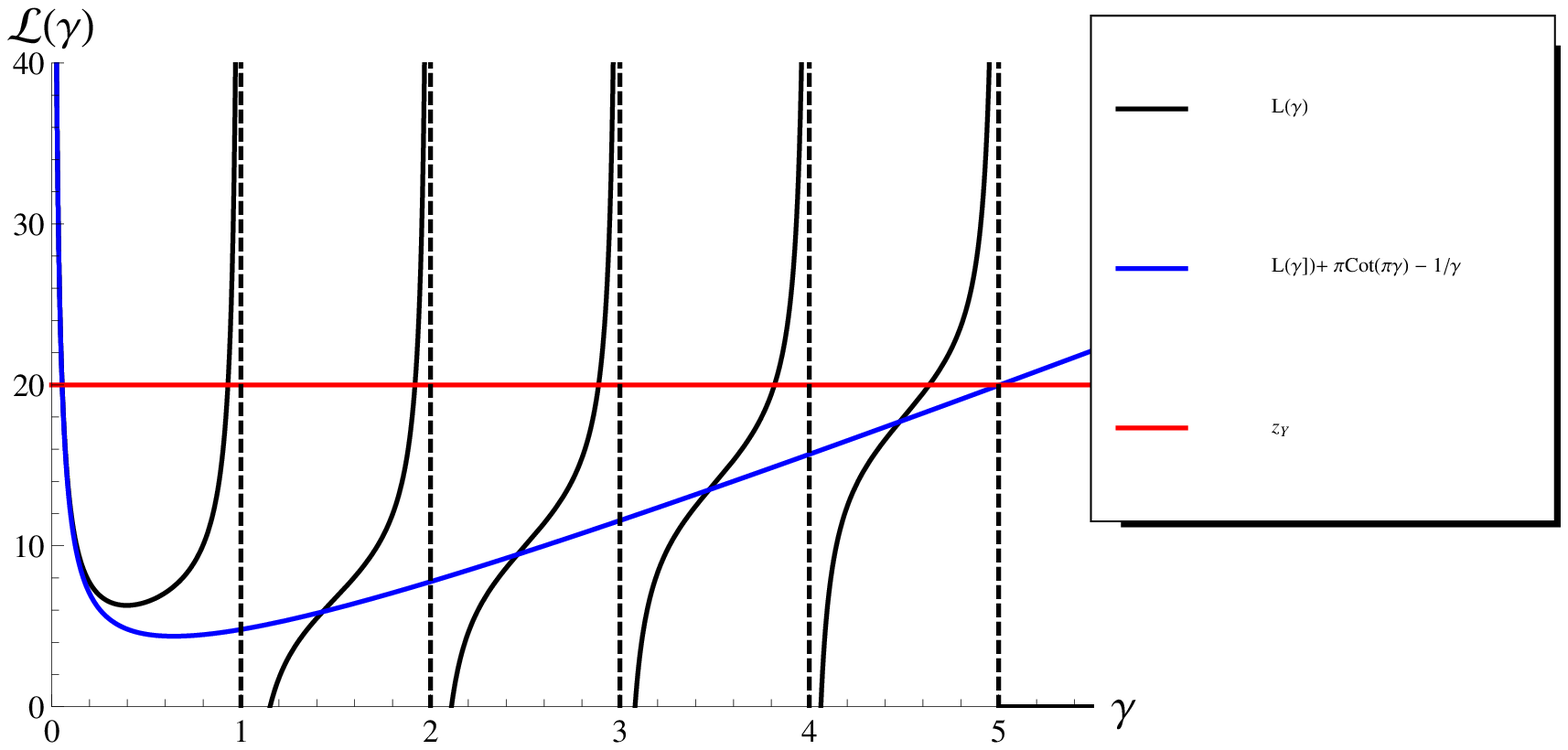,width=180mm, height=65mm}\\
\end{tabular}
\caption{Function ${\cal L}\Lb \ga = d \phi\Lb z\Rb/ d z\Rb $  versus $\ga$ (black line). Blue curve shows ${\cal L}\Lb \ga \Rb + \pi \cot\Lb \pi  \ga \Rb - 1/\ga$ while the red one shows ${\cal L}\Lb \ga\Rb = z_Y$.}
\label{l}
}
\begin{enumerate}
\item \quad  For $z_Y\,>\,z_Y^{\mbox{min}}\,\approx 6.3$ there is a solution at $\phi'_z\Lb z \Rb \,=\ga \Lb z \Rb\,\ll\,1 $ ( $\ga \to 0$);

\item \quad For all $z_Y$ we have solutions $ \phi'\Lb z \Rb\,\,=\,\,\,n \,z $ ($ \gamma \to n$), where $n \,=\,1,2,3,\dots$;

\item \quad For $z_Y\,>\,z_Y^{\mbox{min}}\,\approx 6.3$ we have the following solution:
\beq \label{AS2}
\phi\Lb z\Rb\,\,=\,\,\frac{1 - \ga_{cr}}{\chi\Lb \ga_{cr}\Rb}\,z_Y \,z
\eeq
\end{enumerate} 
In \fig{l} the blue solid  line shows 
\beq \label{LTILDA}
\tilde{\cal L}\,\,\equiv\,\,{\cal L}\Lb \ga\Rb + \pi \cot \Lb \pi \ga\Rb - 1/\ga.
\eeq All poles at $\ga = n$ are excluded in this function while the behaviour at large $\ga$ this function has the same as ${\cal L}\Lb \ga\Rb$. 
This asymptotic behaviour at $\gamma$ close to $z_{Y}/C\Lb\ga_{cr}\Rb$   can be translated in \eq{AS2}  (see also below  \eq{LT1} and  \eq{LT3}).

Therefore, one can see that we face two major problems in searching the solution: (i) at first sight we have infinite number of solutions even in this simplified case; and (ii) we need to find the solution if we exclude the singularities in ${\cal L}\Lb \ga \Rb$.

The infinite number of solutions contradicts the common sense intuition that the physical problem has the only one solution being formulated correctly.  In the next section 3 we show   mathematical arguments that will discriminate different solutions and will select the only one solution. This physical solution will be found  in  section 4.

\begin{boldmath}
\section{Semiclassical solution for $\phi\Lb z, Y'\Rb$}
\end{boldmath}
\subsection{Equations}

For large $\phi\Lb z, Y'\Rb$ \eq{EQD} can be re-written in the form\footnote{We will denote below  by $z_Y$ the sum $z_Y\, +\, 4\,N_0$ and hope it will not cause any difficulties in understanding.}
\beq \label{EQAS}
\tilde{\om}\,\,+\,\,{\cal L}\Lb \ga \Rb\,\,-\,\,{\cal F}\Lb \gamma\Rb e^{ - \phi}\,\,
-\,\,z_Y\,\,=\,\,\tilde{\om}\,\,+\,\,{\cal G}\Lb \ga, \phi \Rb\,\,-\,\,z_Y\,\,=\,\,0
\eeq
 \FIGURE[ht]{
\epsfig{file=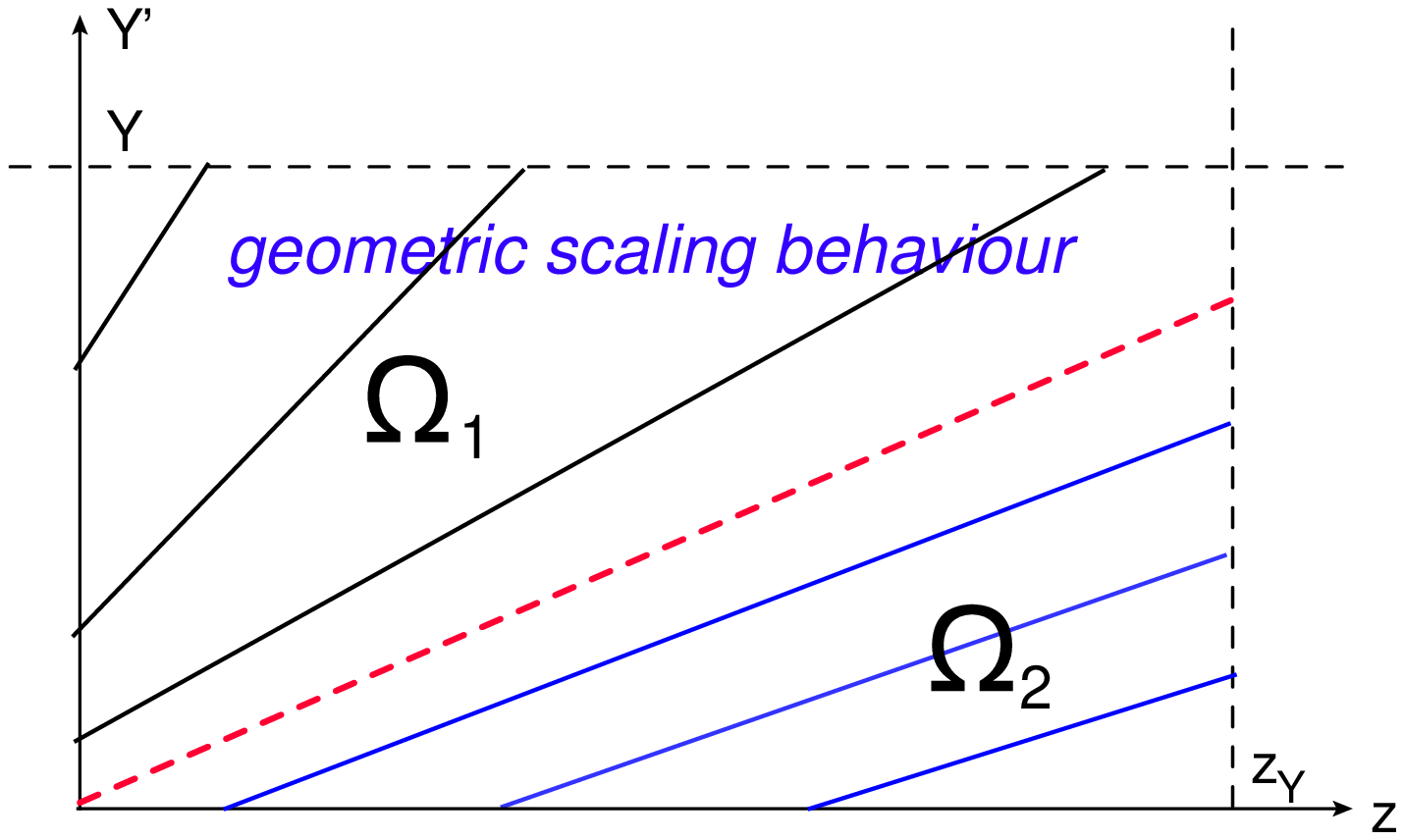,width=65mm}
\caption{ Two kinematic regions: $\Omega_1$ and $\Omega_2$. The solid lines describe  two sets of trajectories in $\Omega_1$ and $\Omega_2$. The trajectories in $\Omega_1$ start at  $z=0$ with given value of $\phi_0$. The trajectories in $\Omega_2$ start at any point of $z$ at $Y'=0$. The initial value of $\ga_0 = \phi_0\,\exp\Lb z_0\Rb$.
The dotted red line  is  the common characteristic line for both $\Omega_1$ and $\Omega_2$. It  has the form (see \eq{OM1} and \eq{OM2} below): $ Y' = z/C\Lb \ga_{cr}\Rb$ . }
\label{2re}}  

We solve this equation in semi-classical approximation assuming that $\phi\Lb z, Y'\Rb$ is a smooth function of both variables $z$ and $Y'$. It is known (see Refs. \cite{HART} and references therein)
that for the equation in the form
\beq \label{SC2}
F(Y^{\prime},z, \phi,\,\gamma,\tilde{\omega})=0
\eeq
with smooth $\phi$. We can introduce the set
of
characteristic lines :  $z(t), Y^{\prime}(t), \phi(t),$ $ \tilde{\omega}(t),$ and $
\gamma(t)$ which are the
functions of the variable $t$ ( artificial time), that satisfy the following
equations:
\begin{eqnarray}
&&\hspace{-0.3cm}(1.)\,\,\,\,\frac{d z}{d\,t}\,\,=\,\,F_{\gamma}\,\,= \,\, \frac{\D {\cal G}\Lb \ga,\phi\Rb}{\D \ga}\nn\\
\,\,\,\,\,\,\,\,\,\,\,\,&&\hspace{-0.3cm}(2.)\,\,\,\,\,\,\,
\frac{d\,Y^\prime}{d\,t}\,\,=\,\,F_{\tilde{\omega}}\,\,=\,\,1\,\,\,\,\,\,\,\nn\\
 &&\hspace{-0.3cm}(3.)\,\,\,\,
\frac{d\,\tilde{\phi}}{d\,t}\,\,=\,\,\gamma\,F_{\gamma}\,+\,\tilde{\omega}\,F_{\tilde{\omega}}\,\,=\,\,  \,\,\ga \frac{\D {\cal G}}{\D \ga}\,\,+\,\,\tilde{\om}
\nonumber \\
&&\hspace{-0.3cm}(4.)\,\,\,\,\frac{d\,\gamma}{d\,t}\,\,=\,\,-
(\,F_{z}\,+\,\gamma\,F_{\phi}\,)\,\,=\,\,-\,\ga\,{\cal F}\Lb \ga \Rb \,e^{ - \phi}\nn\\
&&\hspace{-0.3cm} (5.)\,\,\,\,
\frac{d\,\tilde{\omega}}{d\,t}\,\,=\,\,- \,(\,F_{Y^{\prime}}  \,+\,\tilde{\omega}\,F_\phi\,)\,\,=\,\,-\,\tilde{\om}\,{\cal F}\Lb \ga \Rb \,e^{ - \phi}\label{S}
\end{eqnarray}

From \eq{S}-2 one can see that we can introduce $t = Y'$. Taking the ration of \eq{S}-4 to \eq{S}-5 we obtain that
\beq \label{GAOM}
\tilde{\om}\Lb Y'\Rb\,\,=\,\,{\cal K}\Lb \ga_0\Rb \,\ga\Lb Y'\Rb
\eeq
where $\ga_0 \,\equiv\,\ga\Lb Y' = 0\Rb $ is the initial value of $\ga$ which has to be found from the initial conditions.\

Plugging \eq{GAOM} into \eq{EQAS} we reduce the system of equation (see \eq{S}) to the following set of the equations:
\bea 
&&e^{-\phi}\,\,=\,\,\Big( {\cal L}\Lb \ga\Rb \,\,-\,\,z_Y\,\, +\,\,\ga\,{\cal K}\Lb \ga_0\Rb\Big){\Big/} {\cal F}\Lb \ga \Rb;\label{S21}\\
&&d z\Lb Y'\Rb/d Y' \,\,=\,\,{\cal L}_\ga\Lb \ga \Rb\,\,-\,\,{\cal F}_\ga\Lb \ga \Rb\,e^{-\phi};\label{S22}\\
&&d \ga /d Y'\,=\,-\,\ga\,\Big( {\cal L}\Lb \ga \Rb + {\cal K}\Lb \ga_0\Rb\,\ga - z_Y\Big) \equiv  T\Lb \ga, \ga_0 \Rb;\label{S23}
\eea

\vspace{0.8cm}
It turns out the  general set of trajectories can be divided in two groups: the trajectories that start  at $z =0$ and at arbitrary $Y'$ ( the vertical axis in \fig{2re}) and the trajectories which starting points lie on the horizontal axis in \fig{2re} at $Y'$ and arbitrary $z$. We need to consider these two sets separately.~

\begin{boldmath}
\subsection{Solutions in $\Omega_1$}
\end{boldmath}
In $\Omega_1$  we need to use the initial condition $\phi\Lb z = 0 \Rb = \phi_0$ to obtain the equation for $
{\cal K}\Lb \ga_0\Rb$. 
\beq \label{K1}
{\cal K}\Lb \ga_0\Rb\,\,=\,\,\frac{1}{\ga_0}\Big( -\,{\cal L}\Lb \ga_0\Rb\,\,+\,\,{\cal F}\Lb \ga_0\Rb\,e^{-\phi_0}\,\,+\,\,z_Y\Big)
\eeq

Substituting \eq{K1} into \eq{S21}-\eq{S23} we have the final set of equations for the trajectories
in the region $\Omega_1$.
 \FIGURE[th]{
\begin{minipage}{8cm}{
\epsfig{file=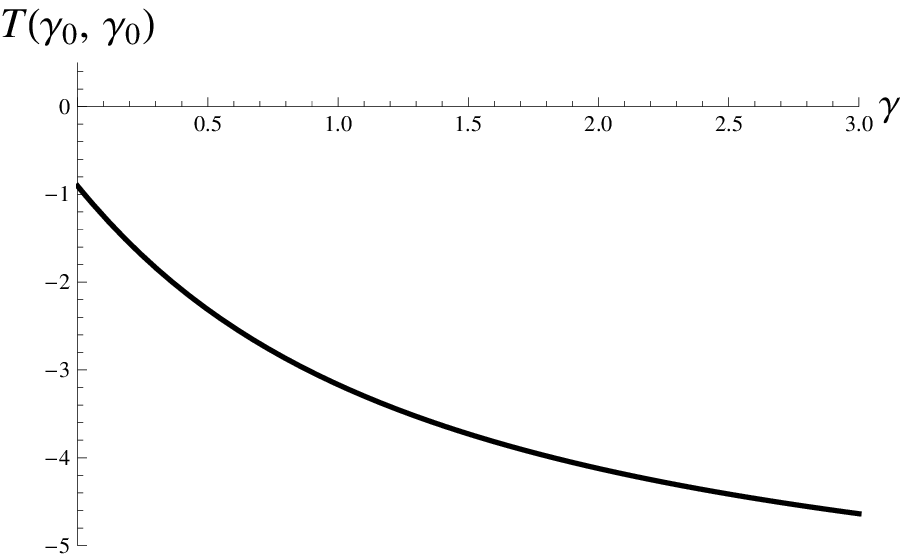,width=75mm}}
\end{minipage}
\begin{minipage}{7cm}{
\caption{$T\Lb \ga_0,\ga_0\Rb$ versus $\ga_0$ in the region $\Omega_1$ . $N_0 = 0.1$ and  $z_Y = 200$.}\label{t00}}
\end{minipage}
}  

In  \fig{t00} we see that function ${\cal T}\Lb \ga_0,\ga_0\Rb$ is negative for all values of $\ga_0$. It means that $\ga\Lb Y'\Rb$ falls down and becomes frozen at the value $\bar{\ga}_0\,=\,\ga_{0,1}\Lb \ga_0\Rb$ at which $T\Lb \bar{\ga}_0,\ga_0\Rb \,=\,0$ (see \fig{t}). One can see that for $0 \,<\,\ga\,<\,1$ actually equation $T\Lb \bar{\ga}_0,\ga_0\Rb \,=\,0$ has two solutions $\ga_{0,1}$ and $\ga_{0,2}$ but both are smaller than $\ga_0$.  For $\ga \,>\,1$ we have the only one solution $\bar{\ga}_0 \,<\,\ga_{0}$. Actually, $\bar{\ga}_0 $ is very close to $\ga_0$. In other words, $\ga$ decreases very fast to $\bar{\ga}_0$. At $\ga = \bar{\ga}_0$ $T\Lb \ga = \bar{\ga}_0, \ga_0\Rb\,=\,0$.
Since the scattering amplitude $N$ should be less that unity from the $s$-channel unitarity we expect that $\exp\Lb - \phi\Lb z, Y'\Rb\Rb$   in entire kinematic region should be positive and not equal to zero.
 \FIGURE[ht]{
\begin{tabular}{c c c }
\epsfig{file=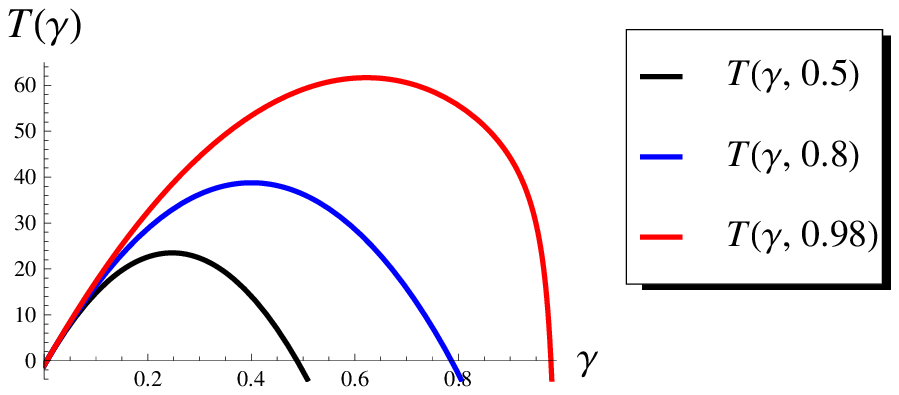,width=55mm}&\epsfig{file=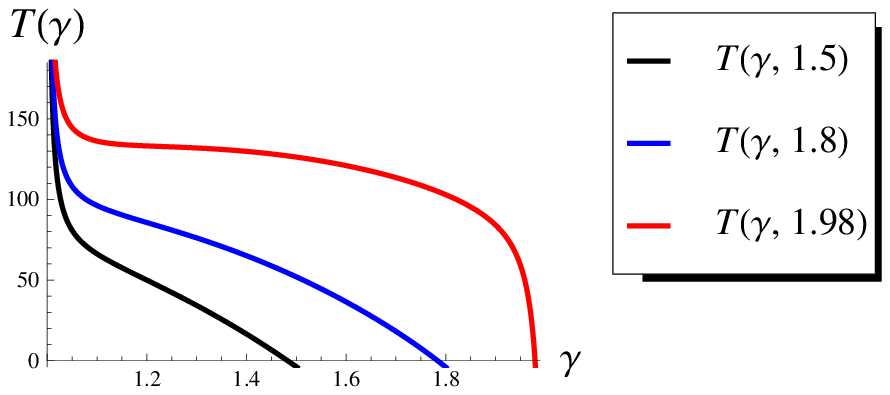,width=55mm}&\epsfig{file=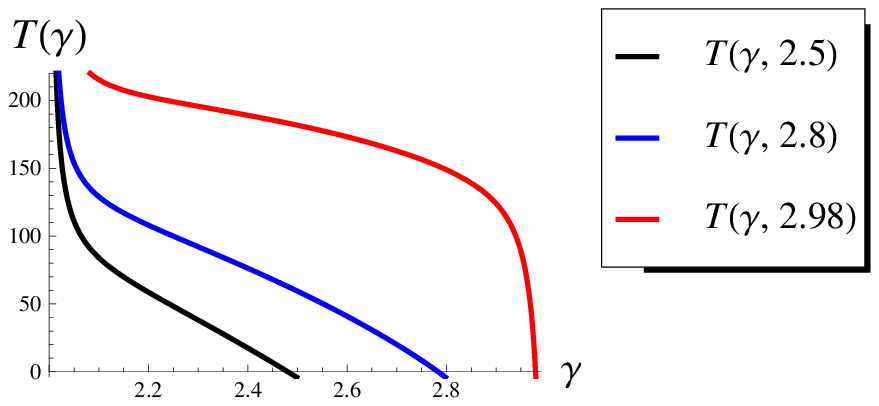,width=55mm}\\
\fig{t}-a & \fig{t}-b & \fig{t}-c\\
\end{tabular}
\caption{$T\Lb \ga,\ga_0\Rb$ versus $\ga$ for $\ga_0= 0 \dots 1$(\fig{t}-a), for $\ga = 1 \dots 2$( \fig{t}-b) and $\ga = 2 \dots 3$(\fig{t}-c) in the region $\Omega_1$. $N_0 = 0.1$ and  $z_Y = 200$.}
\label{t}
}  
\fig{b} shows that we have a different behaviour of the solution for $\exp\Lb - \phi\big( z\Lb Y'\Rb\Rb\big)$ on the trajectories, which contradicts our expectation from the physics point of view.

Therefore, we can conclude that the semi-classical solution that satisfies the physical criterion, does not exist at any $\ga_0$ in the region $\Omega_1$  when
 $\gamma$  close to $n$  with $n = 0,1,2,3, \dots$.

 \FIGURE[ht]{
\begin{tabular}{c c c }
\epsfig{file=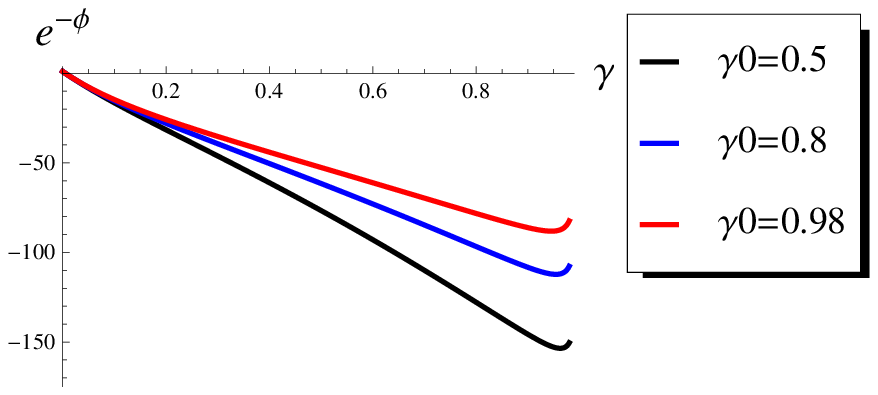,width=55mm}&\epsfig{file=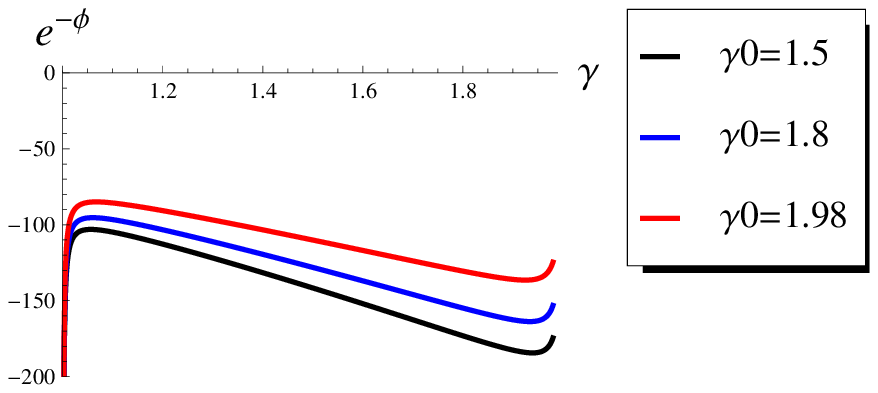,width=55mm}&\epsfig{file=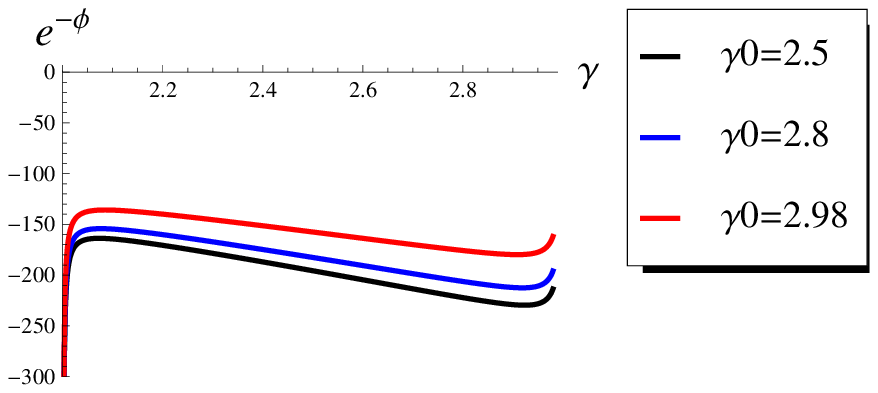,width=55mm}\\
\fig{b}-a & \fig{b}-b & \fig{b}-c\\
\end{tabular}
\caption{$\exp\Lb - \phi\Lb \ga,\ga_0\Rb\Rb$ versus $\ga$ for $\ga_0= 0 \dots 1$(\fig{b}-a), for $\ga = 1 \dots 2$( \fig{b}-b) and $\ga = 2 \dots 3$(\fig{b}-c) in the region $\Omega_1$. $N_0 = 0.1$ and  $z_Y = 200$.}
\label{b}
}  

\begin{boldmath}
\subsection{Solutions in $\Omega_2$}
\end{boldmath}

In the region $\Omega_2$ we have to repeat our analysis since the equation for  $
{\cal K}\Lb \ga_0\Rb$ should be based on the initial condition: $\phi\Lb z=z_0,Y'=0\Rb\,=\,\phi_0 \exp\Lb z_0 \Rb$. Since $\ga_0 =\phi\Lb z=z_0,Y'=0\Rb$ for this initial condition \eq{K1} can be re-written in the form

\beq \label{K2}
{\cal K}\Lb \ga_0\Rb\,\,=\,\,\frac{1}{\ga_0}\Big( -\,{\cal L}\Lb \ga_0\Rb\,\,+\,\,{\cal F}\Lb \ga_0\Rb\,e^{-\ga_0}\,\,+\,\,z_Y\Big)
\eeq
 \FIGURE[th]{
\begin{minipage}{8cm}{
\epsfig{file=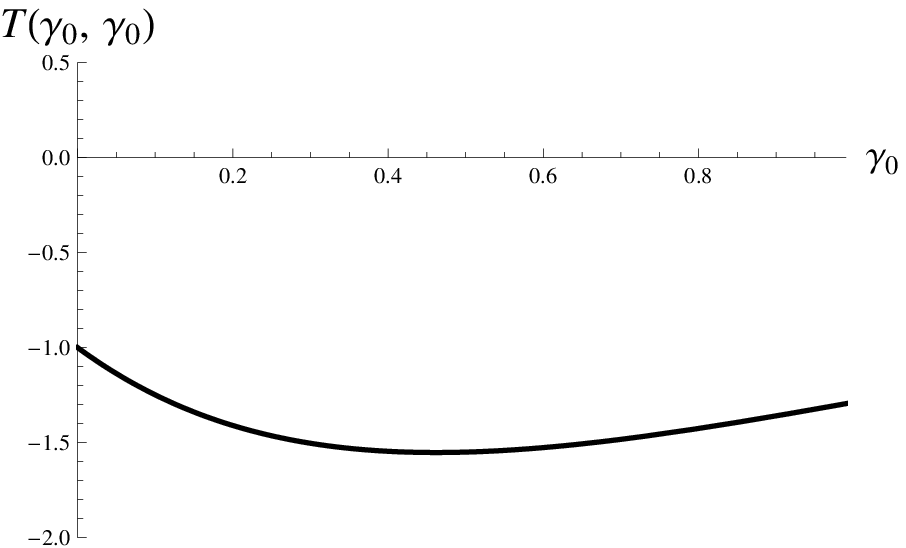,width=75mm}}
\end{minipage}
\begin{minipage}{7cm}{
\caption{$T\Lb \ga_0,\ga_0\Rb$ versus $\ga_0$ in the region $\Omega_2$. $N_0 = 0.1$ and  $z_Y = 200$.}\label{t200}}
\end{minipage}
}  
One can see in \fig{t200} that $T\Lb \ga_0, \ga_0\Rb$ in this region is negative. Therefore, in this region as in region $\Omega_1$ $\ga\Lb z \Rb$ falls down on the trajectory for all trajectories. It turns out that for each value of $\ga_0$ function $T\Lb \ga , \ga_0\Rb$ vanishes at $\bar{\ga}_0 = \ga_{0, 1}\Lb \ga_0\Rb$ ($T\Lb \bar{\ga}_0,\ga_0\Rb = 0$) (see \fig{t2}). Therefore, the solution in region $\Omega_2$ has the same properties as the solution in region $\Omega_1$ (see \fig{t2} and \fig{b2}.

Thus we can repeat the same conclusions for the region $\Omega_2$ as for $\Omega_1$: there is no solutions that satisfy the physical criteria.

 \FIGURE[ht]{
\begin{tabular}{c c c }
\epsfig{file=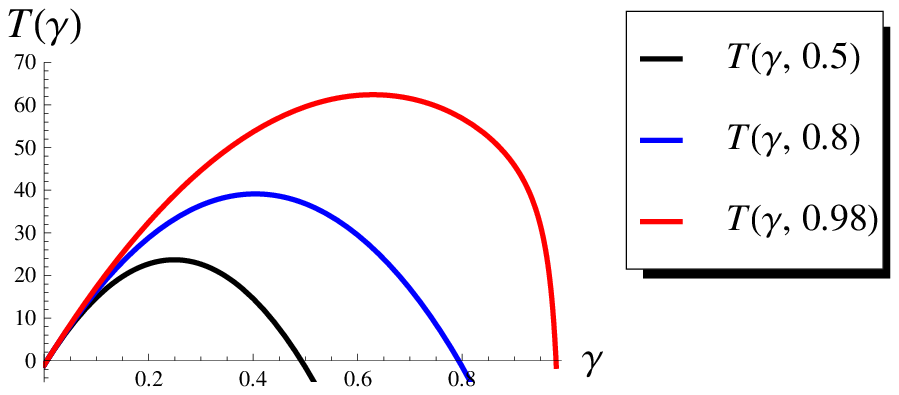,width=55mm}&\epsfig{file=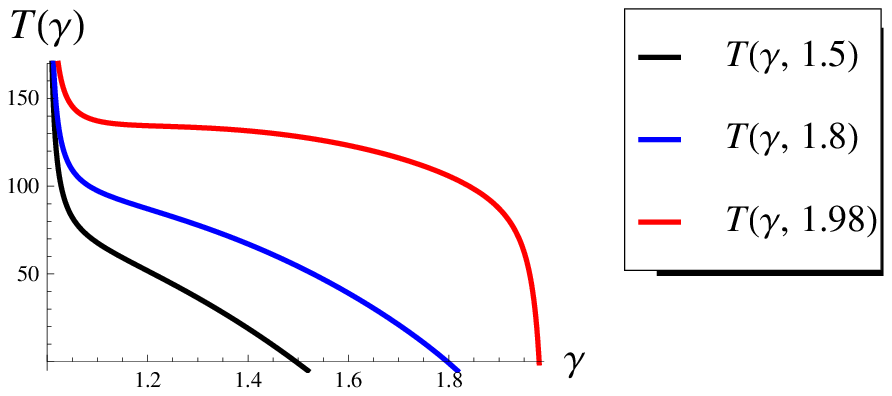,width=55mm}&\epsfig{file=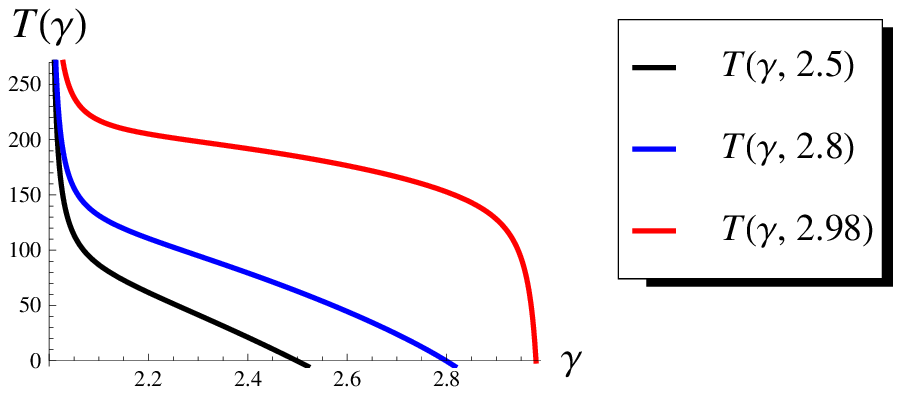,width=55mm}\\
\fig{t2}-a & \fig{t2}-b & \fig{t}-c\\
\end{tabular}
\caption{$T\Lb \ga,\ga_0\Rb$ versus $\ga$ for $\ga_0= 0 \dots 1$(\fig{t2}-a), for $\ga = 1 \dots 2$( \fig{t2}-b) and $\ga = 2 \dots 3$(\fig{t2}-c) in the region $\Omega_2$. $N_0 = 0.1$ and  $z_Y = 200$.}
\label{t2}
}  
 \FIGURE[ht]{
\begin{tabular}{c c c }
\epsfig{file=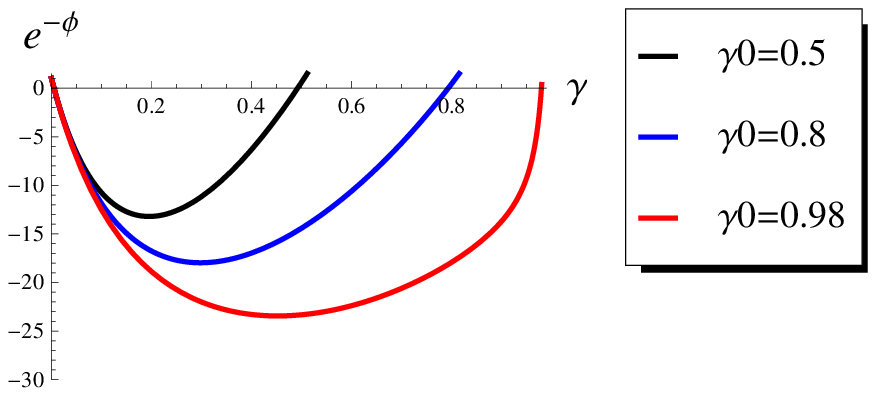,width=55mm}&\epsfig{file=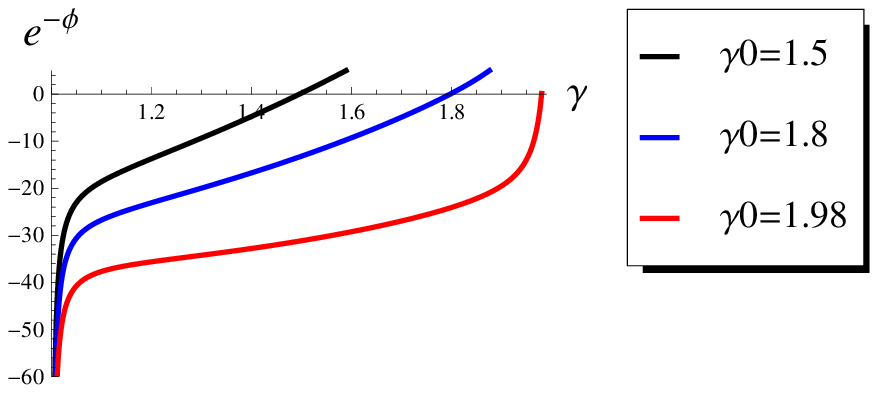,width=55mm}&\epsfig{file=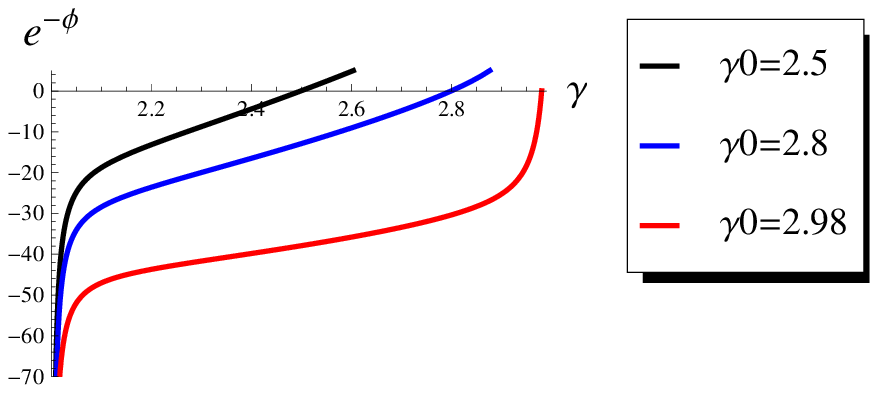,width=55mm}\\
\fig{b2}-a & \fig{b2}-b & \fig{b2}-c\\
\end{tabular}
\caption{$\exp\Lb - \phi\Lb \ga,\ga_0\Rb\Rb$ versus $\ga$ for $\ga_0= 0 \dots 1$(\fig{b2}-a), for $\ga = 1 \dots 2$( \fig{b2}-b) and $\ga = 2 \dots 3$(\fig{b2}-c) in the region $\Omega_2$. $N_0 = 0.1$ and  $z_Y = 200$.}
\label{b2}
}  

     \begin{boldmath}
  
\section{Asymptotic solution with  $\ga \longrightarrow z_Y \Big{/} \Lb \bas \,\frac{\chi\Lb \ga_{cr}\Rb}{1 - \ga_{cr}}\Rb$}

\end{boldmath}


Finally, the only solution which we need to consider is the solution with large $\ga=d \phi\Lb z, Y' \Rb/d z \to z_Y/C\Lb \ga_{cr}\Rb \gg 1$. \eq{EQD} reduces to the simple form
\beq \label{ASEQYZ}
\frac{\partial \phi\Lb z, Y'\Rb}{\partial \bas Y'}\,\,+\,\,C\Lb \ga_{cr}\Rb \frac{\partial \phi\Lb z, Y'\Rb}{\partial z}\,\,=\,\,z_Y~~~~~~~~~~~~\mbox{where}~~~~~~~~~  C\Lb \ga_{cr}\Rb\,\,=\,\,\frac{\chi\Lb \ga_{cr}\Rb}{1 \,-\,\ga_{cr}}
\eeq
with the following initial conditions:
\beq \label{ICAS}
\phi\Lb z\,=\,0,Y'\Rb\,\,=\,\,\phi_0\,;~~~~~~~~~~~~~~~~~~~\phi\Lb z,Y\,=\,0'\Rb\,\,=\,\,\phi_0\,e^z;
\eeq

We  need to consider two separate kinematic regions $\Omega_1$ and $\Omega_2$ for searching the solutions for this equation (see \fig{2re}): $Y' \,\geq \,z/C\Lb \ga_{cr}\Rb$ and $  Y' \,\leq \,z/C\Lb \ga_{cr}\Rb$ \cite{KLT}. In both regions the general solution to \eq{ASEQYZ} takes the form
\beq \label{SOLG}
\phi\Lb z, Y'\Rb\,\,=\,\,\h\, z_Y\,\Lb  Y' \,+\,z/C\Lb \ga_{cr}\Rb \Rb \, \,+\,\,G\Lb Y' \,-\,z/C\Lb \ga_{cr}\Rb \Rb
\eeq
where $G$ is the arbitrary function that has to be found from \eq{ICAS}. In region $\Omega_2$ 
one can see that
\beq \label{OM2}
G_2\,=\,\phi_0\,\exp\Big( - C\Lb \ga_{cr}\Rb\,\Lb Y' \,-\,z/C\Lb \ga_{cr}\Rb\Rb\Big)
\eeq
from the second of \eq{ICAS}. However, in region $\Omega_1$ we need to use the first of \eq{ICAS} and we obtain that
\beq \label{OM1}
G_1\,\,=\,\,-\h\, z_Y\,\Lb  Y' \,-\,z/C\Lb \ga_{cr}\Rb \Rb\,\,+\,\,\phi_0
\eeq

One can see that at $Y' = z/C\Lb \ga_{cr}\Rb$ two $\phi$'s: $\phi_1\Lb z\Rb\,\,=\,\,z_Y\,z\,\,+\,\,\phi_0$ 
and    $\phi_2\Lb z, Y'\Rb\,\,=\,\,\h\, z_Y\,\Lb  Y' \,+\,z/C\Lb \ga_{cr}\Rb \Rb \, \,+\phi_0\,\exp\Big( - C\Lb \ga_{cr}\Rb\,\Lb Y' \,-\,z/C\Lb \ga_{cr}\Rb\Rb\Big)$ are equal, providing the needed matching.

Therefore, the simplified \eq{ASEQYZ}  leads to the geometric scaling solution for $Y' \,\,\geq\,\,z/C\Lb \ga_{cr}\Rb$  while for $Y' \,\,\leq\,\,z/C\Lb \ga_{cr}\Rb$ we have a solution with explicit scaling violation.

Armed with the asymptotic solution given by \eq{OM1} and \eq{OM2} we study  the numerical solution to \eq{EQD}  in region $\Omega_1$ assuming the geometric scaling behaviour and  considering the following iterative procedure. 
  Plugging in \eq{EQD}  $\phi\Lb z, Y'\Rb = \phi\Lb z\Rb$  and  \eq{LTILDA} we see that
this equation takes the following form in this region
\beq \label{LT1}
 \frac{\chi\Lb \ga_{cr}\Rb}{1 - \ga_{cr}} \,\frac{d \phi^{(i)}\Lb z \Rb}{d z}\,\,-\,\,2\Lb \ga_E + \ln\Lb \frac{d \phi^{(i)}\Lb z \Rb}{d z}\Rb\Rb\,\,=\,\, z_Y \,\,\,-\,\,H\Lb z \Rb \,-\,H\Lb z_Y - z\Rb \,\,\,\,\,\mbox{with}\,\,\,\,\,H\Lb z \Rb\,\,=\,\,\int^z_0d z' e^{- \phi^{(i-1)}\Lb z'\Rb}
\eeq
where $\ga_E$ is the Euler constant ($\ga_E\,=\,0.5777216$) and where $\phi^{(i)}$ is the solution of the $i$-th iteration of the equation. Deriving \eq{LT1} we used that $\tilde{\cal L} \,\xrightarrow{ \ga \gg 1}\,\ga_E \,+\,\ln\Lb \ga \Rb$. 

We solve \eq{LT1} using   iteration procedure with 
\beq \label{LT2}
 \phi^{(i=0)}\,=\, \Big\{\Lb z_Y\,+\,2\Lb \ga_E + \ln z_Y\Rb\Rb(1 - \ga_{cr})/\chi\Lb\ga_{cr}\Rb\Big\}\,z.
 \eeq
  The result of the third iteration is shown in \fig{lt}. One can see from \fig{lt}-a that the solution is very close to $\phi^{(0)}$ given by \eq{LT2}. $N'_z$ reaches the unitarity limit at small   values of $z$($z \leq 0.5$).
In \fig{lt}-c we plotted the difference between the l.h.s. and the r.h.s. of \eq{EQD}. This difference can be written in the form:
\bea \label{TI}
&&T^{(i)}\Lb z\Rb \,\,=\\
&&\,\,{\cal L}\Lb \phi'^{(i)}_z\Lb z \Rb\Rb - {\cal F}\Lb \phi'^{(i)}_z\Lb z \Rb \Rb \,e^{ - \phi^{(i)}\Lb z \Rb}\,\,-\,\,\tilde{N}^{(i)}\Lb z \Rb \,\,-\,\,\tilde{N}^{(i)}\Lb z_Y - z \Rb\nn\\
&& \mbox{where}\,\,\,\,e^{ - \phi^{(i)}}\,\,=\,\,1 - d \tilde{N}^{(i)}\Lb z \Rb/d z\,\,\,\,\mbox{and}\,\,\,\,
d \phi^{(i)}/d z \,\,=\,\,\frac{d^2\tilde{N}^{(i)}\Lb z \Rb}{d z^2}/\Lb 1 - d \tilde{N}^{(i)}\Lb z \Rb/d z\Rb\nn
\eea

$T^{(i)}\Lb z\Rb$ turns out to be small ( less than 1) leading to the accuracy of the solution about 2\%.

 \FIGURE[ht]{
\begin{tabular}{c c c }
\epsfig{file=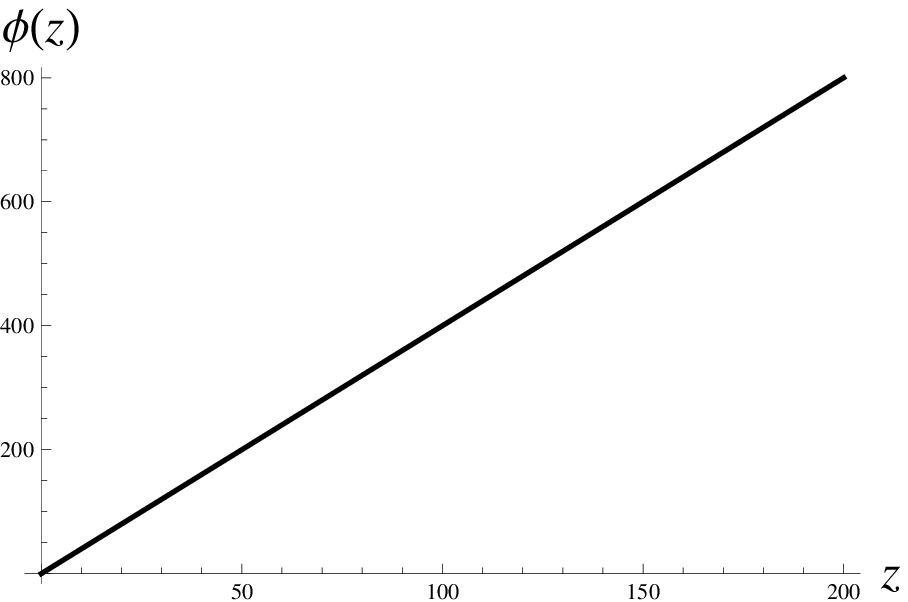,width=55mm}&\epsfig{file=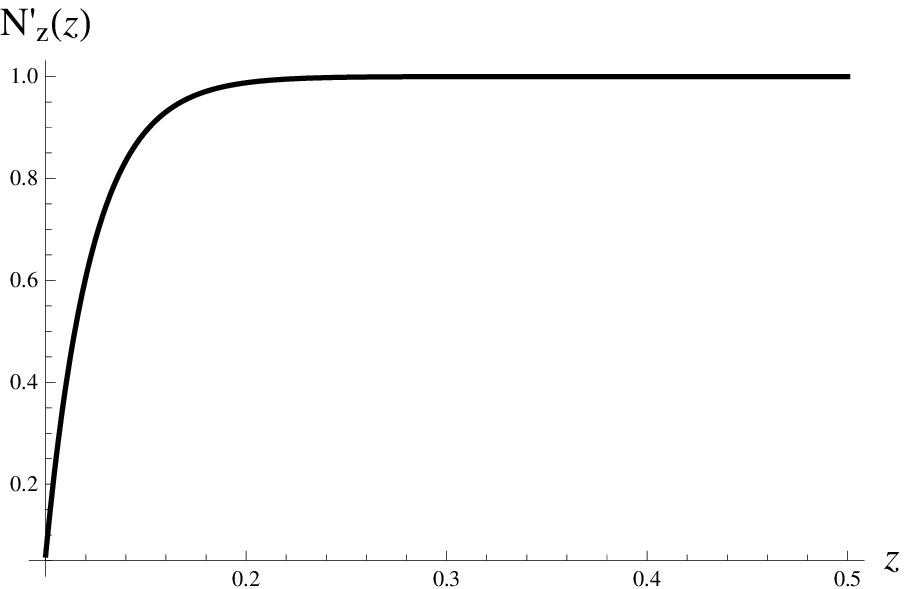,width=55mm}&\epsfig{file=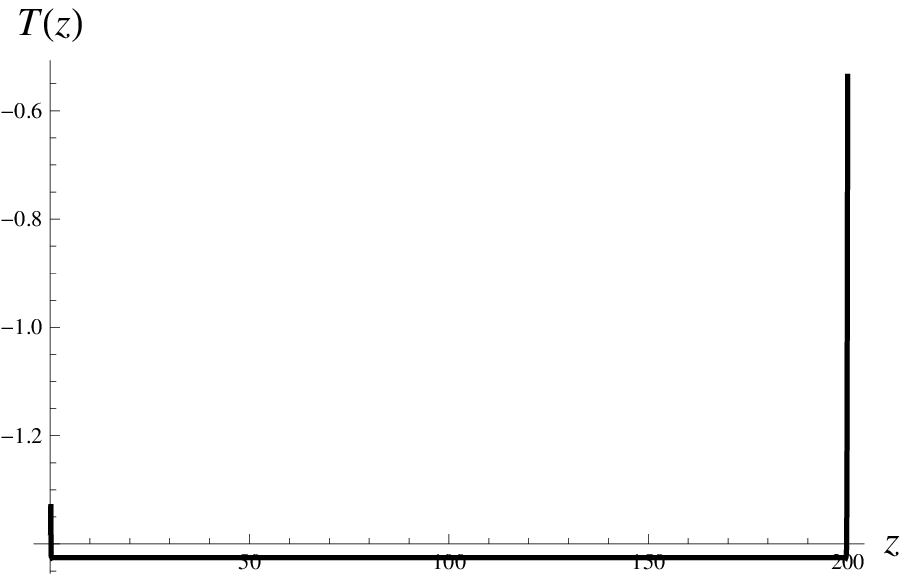,width=55mm}\\
\fig{lt}-a & \fig{lt}-b & \fig{lt}-c\\
\end{tabular}
\caption{The third iteration of \eq{LT1}:  solutions  for  functions $\phi\Lb z\Rb$(\fig{lt}-a) and the scattering amplitude $N'_z\Lb z\Rb$ (\fig{lt}-b) versus $z$. \, In \fig{lt}-c we plot function $T\Lb z \Rb$ (see \eq{TI}). In the figure the following parameters are taken: $N_0 = 0.1$ and  $z_Y = 200$.}
\label{lt}
}  
For the region $\Omega_2$ (see \fig{2re})  we solve a more general equation for  $\phi\Lb z,Y'\Rb$. In this case \eq{LT1} takes the form
\beq \label{LT3}
\frac{\partial \phi\Lb z, Y'\Rb}{\partial \bas Y'}\,\,+\,
\frac{\chi\Lb \ga_{cr}\Rb}{1 - \ga_{cr}} \,\frac{\partial\phi\Lb z, Y' \Rb}{\partial z}\,\,-\,\,2\Lb \ga_E + \ln\Lb \frac{d \phi\Lb z, Y'\Rb}{d z}\Rb\Rb\,\,=\,\, z_Y \,\,+4\,N_0\,\,-\,\,H\Lb z ,Y'\Rb \,-\,H\Lb z_Y - z, Y  - Y'\Rb
\eeq
where  $\,H\Lb z, Y' \Rb\,\,=\,\,\int^z_0d z' e^{- \phi\Lb z', Y'\Rb}$.  The initial conditions are given by \eq{ICAS}.

The numerical solution is shown in \fig{lts}. One can see that at large values of $Y'$  solution approaches the geometric scaling solution of \eq{LT1}.
We can use  this  physical solution, shown in \fig{lts} and  \fig{lt},  to study  the N-N scattering  in the next section.

 \FIGURE[ht]{
\begin{tabular}{c}
\epsfig{file=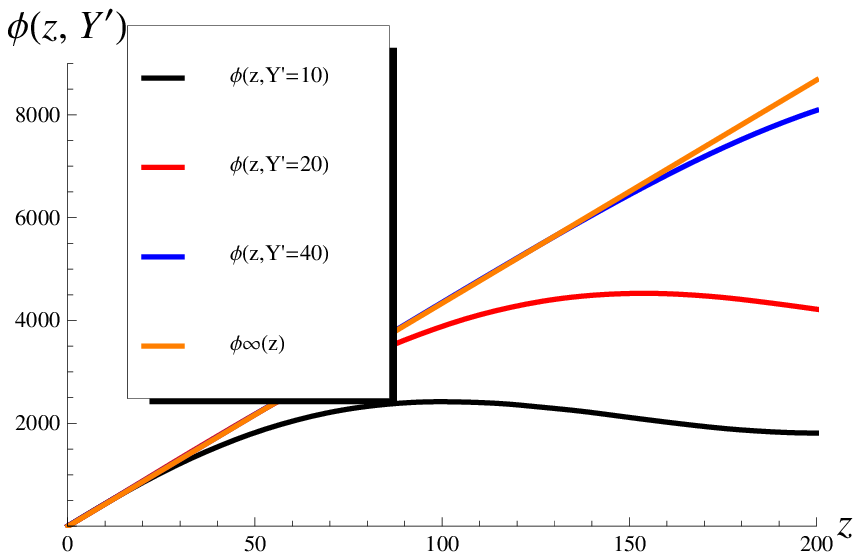,width=100mm}\\
\end{tabular}
\caption{ Function $\phi\Lb z, Y'\Rb$ at different values of $Y'$ as function of $z$.    In the figure the following parameters are taken: $N_0 = 0.1$ and  $z_Y = 200$ }
\label{lts}
}

\section{Nucleus-nucleus scattering amplitude}

We need to specify   term $S_E$ in \eq{BFKLFI} for calculating the scattering amplitude with a nucleus. This term 
 determines the interaction of the BFKL Pomeron with the nucleons of the nucleus and it has been written in Ref.\cite{BRA} in the following form
 \beq \label{SE}
 S_E\,\,=\,\,\int d^2 b\, d^2 k \,\Big( \Phi\Lb k, b, Y'=0\Rb \tau_{A_2}\Lb k, b \Rb\,\,\,+\,\,\, \Phi^
 \dagger\Lb k, b, Y'=Y\Rb \tau_{A_1}\Lb k, b \Rb\Big)
 \eeq
 where 
 \beq \label{TAU}
 \tau_A\Lb k, b\Rb\,\,=\,\,S_A\Lb b \Rb\,\sigma\Lb k \Rb\,\,\,\,\mbox{with}\,\,\,\,\,S_A\Lb b \Rb\,\,=\,\,\int^{+\infty}_{-\infty}\,d \,z\,\rho\Lb z, b\Rb\,\,\,\,\mbox{and}\,\,\,\,
 \sigma\Lb k\Rb\,=\,\int n\Lb k, b\Rb d^2 b
 \eeq
 In \eq{TAU} $\rho$ is the density of the nucleons in the nucleus and $\sigma \Lb k \Rb$ is the cross section (imaginary part of the forward scattering  amplitude) of the dipole with the nucleon at low energy while $ n\Lb k, b\Rb$ is the dipole-nucleon scattering amplitude.
 
 As we have eluded,  in our treatment of nucleus-nucleus interaction we use the Glauber-type approach\cite{Glauber} integrating over all impact parameters of dipole-dipole and dipole-nucleon interaction since they are assumed  to be much smaller than the impact parameters of nucleon-nucleon scattering  (see for example Ref.\cite{LMKS} in which this approach is discussed in details). The latter is of the order of $R_A$, which is larger than the nucleon radius and the sizes of all interacting dipoles.   As we have discussed, we assume that instead of the real nuclei we are dealing with the nuclei that consist of mesons made of heavy quark and antiquarks. For such mesons we can calculate $n\Lb k \Rb$ in the Born Approximation of perturbative QCD. In coordinate space $n \Lb r\Rb = \Lb  2 \pi \as^2 C_F/ N_c\Rb r^2 \ln\Lb R^2/r^2\Rb\,\Theta\Lb R - r \Rb$ where $R$ is the radius of the meson which is of the order of  $ 1/\Lb \as\Lb M_Q\Rb  M_Q\Rb$ where 
 $M_Q$ is the mass of heavy quark.  In the momentum space we have the following $\tau_A\Lb k , b \Rb$
 \beq \label{TAUF}
 \tau_A\Lb k , b \Rb\,\,\,=\,\,\, S_A\Lb b \Rb\,\sigma\Lb k \Rb\,\,=\,\,S_A\Lb b \Rb\,\frac{\as^2 C_F}{\pi N_c }\,\frac{1}{ k^2 \,+\,\as^2\,M^2_Q}
 \eeq
 
 $\tau$ determine the initial conditions for the amplitude $N\Lb k,b,Y'\Rb$ and $N^\dagger\Lb k, b,Y'\Rb$, namely
 \beq \label{INCON}
 N\Lb k, b, Y'=0\Rb\,\,=\,\,\tau_{A_1}\Lb b, k\Rb\,;\,\,\,\,\,\,\,\, N^\dagger\Lb k, b, Y'=Y\Rb\,\,=\,\,\tau_{A_2}\Lb b, k\Rb\,; 
 \eeq
  
  For  simplicity,  we consider the cylindric nuclei for which the $b$ dependence is given by $\Theta\Lb R_A  -   b\Rb$. 
  In this model 
 \beq \label{SMOD}
  S_A\Lb b \Rb\,\,=\,\,2\,\rho_0\,R_A \,\Theta\Lb R_A - b\Rb
  \eeq
  
  In this simple model the entire $b$ dependence of $N\Lb k, b,Y'\Rb$ and $ N^\dagger\Lb k, b, Y'\Rb$ turns out to be the same as in initial condition leading to  
  \bea \label{MODB}
  N\Lb k, b, Y'\Rb \,\,&=&\,\, S_A\Lb b \Rb \,{\cal \sigma} \Lb k, Y'\Rb  \,\,=\,\, S_A\Lb b \Rb \,\frac{1}{k^2_0}{\cal N} \Lb k, Y'\Rb  \nn\\ 
   N^\dagger\Lb k, b.Y'\Rb \,\,&=&\,\, S_A\Lb b \Rb \,{\sigma}^\dagger \Lb k, Y'\Rb  \,\,=\,\, S_A\Lb b \Rb \,\frac{1}{k^2_0}{\cal N}^\dagger \Lb k, Y'\Rb   
     \eea
  where $k_0$ is the  typical transverse momentum  in the proton ($k_0 \propto\,\as M_Q$). Functions ${\cal N}$ and  $  {\cal N}^\dagger$ are dimensionless and for them we have the geometric scaling behaviour inside the saturation region \cite{GS} and in the vicinity of the saturation outside of the saturation region\cite{IIM}.

   Using  $\rho\Lb z=0,b=0\Rb = \rho_0 = 0.17 1/fm^3$ and $R_A = 1.2\,A^{1/3} \,fm$ we can estimate the value of $\tau$ at
 $k = 0 $ for the gold: $\tau = 8  \rho_0 \,R_A/( 9 \pi \,M^2_Q)\,\,=\,\,0.32(1/fm^2)/M^2_Q$ . One can see that  $N_0 = 0.1$ which we used for the numerical solution  can be reached even at $k=0$ if $M_Q \approx 0.36 \,GeV$.
 
 Calculating the scattering amplitude,  we need first  to sum all two nuclei reducible diagrams (see \fig{amaa} for example). In such diagrams  we can single out one or more states with two nuclei in the $s$-channel (see, for example, two such states in \fig{amaa}).
  \FIGURE[ht]{
\begin{tabular}{c}
\epsfig{file=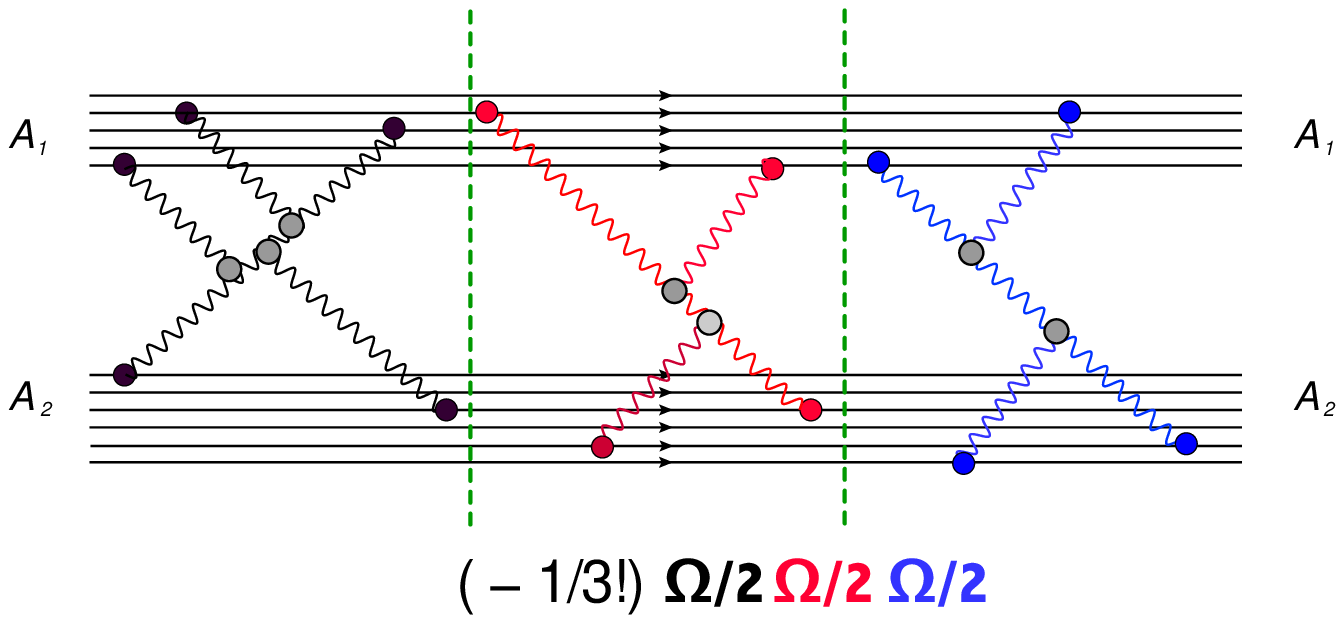,width=100mm}\\
\end{tabular}
\caption{ The example of two nuclei reducible diagram in the scattering amplitude for nucleus-nucleus interaction.}
\label{amaa}
}  
 
 This sum can be written as follows
 \beq \label{AAAM}
 Im A\Lb b, Y\Rb\,\,=\,\,\Big( 1 \,\,\,-\,\,e^{ - \h\Omega\Lb b, Y\Rb}\Big)
 \eeq
The notation : $\Omega $ in \eq{AAAM},  is introduced to be  opacity for the nucleus-nucleus scattering in the Glauber-Gribov approach\cite{Glauber}.
 
 Using \eq{SE} one can see that the equation for the opacity  $\Omega$ has the following form (see \fig{aaeq})
 
\beq \label{OM}
\Omega\Lb b, Y\Rb\,=\,\int d^2 b'\, d^2 k\, \tau_{A_1}\Lb \vec{b} \,-\,\vec{b}', k\Rb \,N\Lb b',k, Y'=Y\Rb\,=\,
 \int d^2 b'\, d^2 k\, N^\dagger\Lb \vec{b} \,-\,\vec{b}', k, Y'=0\Rb \,\tau_{A_2}\Lb b',k\Rb
 \eeq
   \FIGURE[ht]{
\begin{tabular}{c}
\epsfig{file=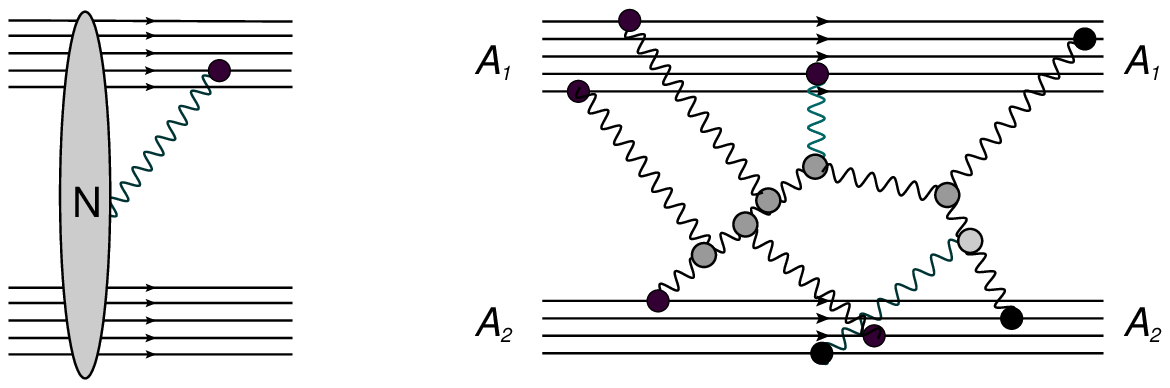,width=100mm}\\
\end{tabular}
\caption{ The equation for opacity $\Omega$.}
\label{aaeq}
}   
 Plugging in \eq{MODB} we can rewrite \eq{OM} in the form
 \beq \label{OMMOD}
 \Omega\Lb b, Y\Rb\,=\,T_{AA}\Lb b \Rb \int \, d^2 k\, \sigma\Lb \vec{b} \,-\,\vec{b}', k\Rb \,{\sigma}\Lb k, Y'=Y\Rb\,\,=\,\,T_{AA}\Lb b \Rb\,\frac{1}{k^2_0} \int \, d^2 k\, \sigma\Lb \vec{b} \,-\,\vec{b}', k\Rb \,{\cal N}\Lb k, Y'=Y\Rb  \eeq 
  where
\bea \label{TAA}
T_{AA}\Lb b\Rb\,\,&=&\Lb 2 \rho_0 R_A\Rb^2\,\,\int d^2 b' \Theta\Lb b' - R_A\Rb \Theta\Lb |\vec{b} - \vec{b}'| - R_A\Rb\,\,\\
&=&\,\,\Lb 2 \rho_0 R_A\Rb^2\, R^2_A\Lb 2\, \arccos\Lb\frac{ b}{2\,R_A}\Rb - \frac{b}{2 \,R_A}\sqrt{ 1 - \frac{ b^2}{4\,R^2_A}}\,\,\Rb\nn
\eea

 As we have discussed the solution for ${\cal N}\Lb z \Rb $  at $z \leq z_{min}$ and $ {\cal N}^\dagger\Lb Z_y - z\Rb $ for 
 $z_Y - z \leq z_{min}$  are   the solution to the linear BFKL equation  which can be written in the following form
  \bea
0\,\leq\, z \,\leq z_{min} \,\,\,\,\,\,\,\,\,&N\Lb k, Y\Rb = \frac{1}{k^2_0}{\cal N}\Lb z \Rb & =\,\, \int \, d^2 k'\, \sigma\Lb  k'\Rb \,{\sigma}_{BFKL}\Lb k/k', Y'=Y\Rb\label{OM10} \\
    & & =\,\,\frac{\as^2 C_F}{\pi \,N_c} \frac{1}{k^2_0}  \int \frac{ d k'^2}{k'^2 \,+\,\as^2 M^2_Q}\,{\cal N}_{BFKL}\Lb k/k', Y'=Y\Rb   \nn\\
 0\,\leq\, z_Y -  z  \,\leq \,z_{min} \,\,\,\,\,\,\,\,\,&N^\dagger\Lb k, Y\Rb \,=\, \frac{1}{k^2_0}{\cal N}^{\dagger} \Lb z \Rb & =\,\, \int \, d^2 k'\, \sigma\Lb  k' \Rb \,{\cal \sigma}^{\dagger}_{BFKL}\Lb k/k', Y'=Y\Rb\label{OM20} \\
    & & =\frac{\as^2 C_F}{\pi \,N_c} \frac{1}{k^2_0}  \int \frac{ d k'^2}{k'^2 \,+\,\as^2 M^2_Q}\,{\cal N}^\dagger_{BFKL}\Lb k/k', Y' = Y\Rb  \nn   
    \eea 
    
    In the vicinity of the saturation scale  ${\cal N}_{BFKL}$ and   ${\cal N}^\dagger_{BFKL}$  in \eq{OM10} and \eq{OM20} can be written in the simple form \cite{GS}:
    
   \beq  \label{OM12}
   {\cal N}_{BFKL}\Lb k/k', Y'=Y\Rb\,\,=\,\,N_0\,e^{\Lb 1 - \gamma_{cr}\Rb z' }\,\,\,\,\,\mbox{and}   \,\,\,\,\   {\cal N}^\dagger_{BFKL}\Lb k/k', Y'=Y\Rb\,\,=\,\,N_0\,e^{\Lb 1 - \gamma_{cr}\Rb\Lb z_Y -  z' 
   \Rb}  
  \eeq
 where 
\beq \label{QS1}
z' \,\,=\,\,\bas\,\frac{\chi\Lb \ga_{cr}\Rb}{1 - \ga_{cr}}\,Y\,\,-\,\,\ln\Lb k^2/k'^2\Rb\,
\eeq
     
  Taking the integration over $k'$ in \eq{OM10} and \eq{OM20} we obtain:
    \bea
0\,\leq\, z \,\leq z_{min} \,\,\,\,\,\,\,\,\,&{\cal N}\Lb z \Rb & =\,\, \frac{\as^2 C_F}{\pi \,N_c} \frac{N_0}{1 - \ga_{cr}} \exp\Big( \Lb 1 - \ga_{cr}\Rb z\Big)\label{OM21}  \nn\\
 0\,\leq\, z_Y -  z  \,\leq \,z_{min} \,\,\,\,\,\,\,\,\,&{\cal N}^{\dagger} \Lb z \Rb & =\,\,\, \frac{\as^2 C_F}{\pi \,N_c} \frac{N_0}{1 - \ga_{cr}} \exp\Big( \Lb 1 - \ga_{cr}\Rb \Lb z_Y -  z\Rb\Big)\label{OM22}      \eea      
     
     It should be noticed that in the above equations  $l$ is defined as $l = \ln\Lb k^2/(\as M_Q)^2\Rb$ (see \eq{QS} and \eq{NN}) with $k^2_0 = (\as M_Q)^2$.

 Having these equation in mind we use a generalization of \eq{OMMOD}
 \beq \label{OMMODG}
\Omega\Lb  b; Y\Rb\,\,=\,\,T_{AA}\Lb b \Rb \int d l \,\,{\cal  N}^\dag \Lb  Y - Y' ; L - l \Rb \,{\cal N}\Lb Y'; l   \Rb
\eeq

\eq{OMMODG} has been proposed and discussed in Ref.\cite{MUSH} and for the nucleus-nucleus scattering it is illustrated by \fig{aa}. From this figure one can see that arbitrary BFKL Pomeron diagram for the case of nucleus-nucleus scattering can be written as the product of $N^\dag N$. The extra Pomeron contribution that could connect two sets of diagrams for $N^\dag$ and $N$(shown in black for $N$ and in red for $N^\dag$ in \fig{aa}) leads to a small corrections (see Ref.\cite{KLM} for details). $Y'$ is the arbitrary rapidity which  is chosen from the condition $ z = z_{min}$.

 \FIGURE[ht]{
\begin{tabular}{c}
\epsfig{file=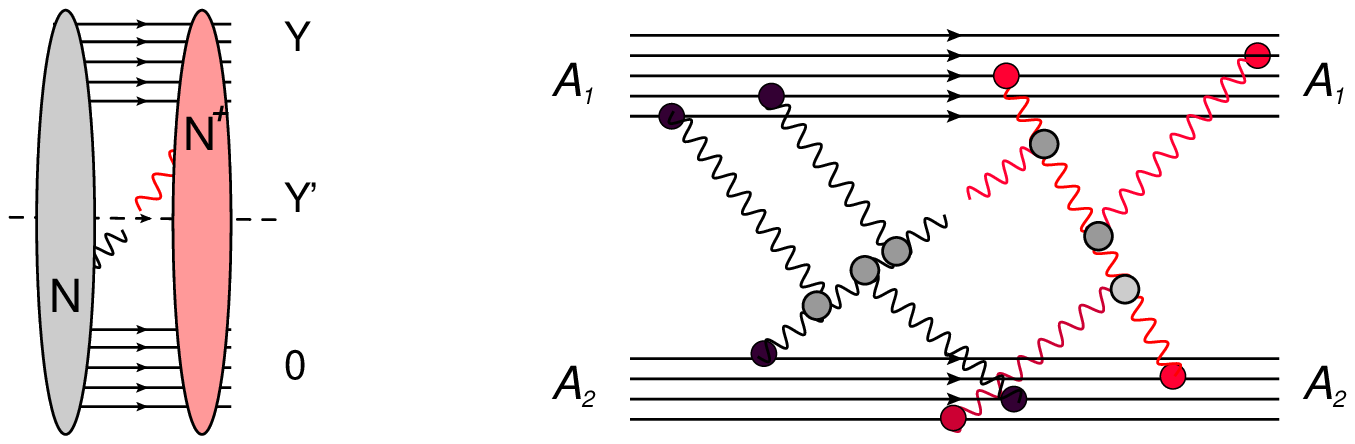,width=100mm}\\
\end{tabular}
\caption{ Graphic form of \eq{OMMODG}. Solid black lines denote the Pomerons and their interactions that contribute to $N$, while the same contributions for $N^\dag$ are shown in  red.}
\label{aa}
}   

Using \eq{OM2} for ${\cal  N}^\dag \Lb  Y - Y' ; L - l \Rb$  we can reduce \eq{OMMODG} to the form
\beq \label{OMF}
 \Omega\Lb b, z_Y\Rb\,=\,T_{AA}\Lb b \Rb \,\frac{\as^2 C_F}{\pi \,N_c\,\Lb \as M_Q\Rb^2}  \frac{N_0}{1 - \gamma_{cr}}\,\int_{z_Y + z_{min}} d z\,e^{ \Lb 1 - \gamma_{cr}\Rb\Lb z_Y - z\Rb} \,{\cal N}\Lb z   \Rb 
 \eeq
 
  This equation stems from the geometric scaling behaviour of the solution to the  equations for $N$ and $N^\dagger$.
  
  In the kinematic region $z \leq 2 z_{min} $ both ${\cal N}$ and ${\cal N}^\dagger$ in \eq{OMMODG}
  are the solution of the BFKL equation and $\Omega$ in this region can be written as follows using \eq{UN}:
  
 \beq \label{OM4}
 z_Y \,\,\leq\, 2 \,z_{min}:\,\,\,\,\,\, \Omega\Lb b, z_Y\Rb\,=\,\frac{T_{AA}\Lb b \Rb }{\Lb \as M_Q\Rb^2}\,\Big(\frac{\as^2 C_F}{\pi \,N_c} \Big)^2\frac{  N^2_0  }{\Lb1 - \ga_{cr}\Rb^2} \exp\Lb\Lb 1 - \ga_{cr}\Rb z _Y \Rb \eeq
 
 For large $ z \,>\,2\, z_{min}$  $N(z)$ is given by solution of \eq{LT1}. For estimates we can use for $N\Lb z \Rb$ in this region the solution in the form
 \beq \label{OM5}
{\cal N}_{z > z_{min}}\Lb z\Rb\,\,=\,\, \int^z_{z_{min}} d \,z' \Big( 1\,\,-\,\,e^{-\phi^{(i = 0)}\Lb z'\Rb}\Big)  \,\,+\,\, {\cal N}\Lb z_{min}\Rb
\eeq
where $\phi^{(i = 0)}$ is given by \eq{LT2}.
The expression for $\Omega$  reads as follows
\bea \label{OM6}
 z_Y \,\,\geq\,\,2\,z_{min}:&\,\,\,\,\,\,& \Omega\Lb b, z_Y\Rb\,=\,\frac{T_{AA}}{\,\Lb \as M_Q\Rb^2}\Big(\frac{\as^2 C_F}{\pi \,N_c} \Big)^2\,\frac{  N_0 }{1 - \ga_{cr}}\Big\{  \int^{z_Y - z_{min}}_{z_{min}} d\,z
 \int^z_{z_{min}} d \,z' \Big( 1\,\,-\,\,e^{-\phi^{(i = 0)}\Lb z'\Rb}\Big)\nn\\
    &\,\,\,\,\,\,& + \frac{N_0  }{(1 - \ga_{cr})} \exp\Big(2\, (1 - \ga_{cr}) \,z _{min} \Big)\Big\}
     \eea

  The result of our estimates using \eq{OM4} - \eq{OM6} is plotted in \fig{ampaa}. One can see that starting with small values of $z_Y \geq 8 $ the amplitude at $b=0$ is close to the unitarity bound
 (see red line in \fig{amaa}).
  For comparison we plot in \fig{ampaa} also the amplitude which  corresponds to the the exchange by one BFKL Pomeron in nucleon-nucleon scattering:
\beq \label{AA9}
A_P\Lb AA; z_Y, b \Rb\,\,=\,\,i\,\Big( 1 \,\,-\,\,\exp\Lb  - \h \Omega_P\Lb z_Y,b\Rb \Rb \Big)\eeq
    with $\Omega_P\Lb z_y,b\Rb$ given by \eq{OM10} at any value of $z_Y$.      
      \FIGURE[ht]{
\begin{tabular}{c}
\epsfig{file=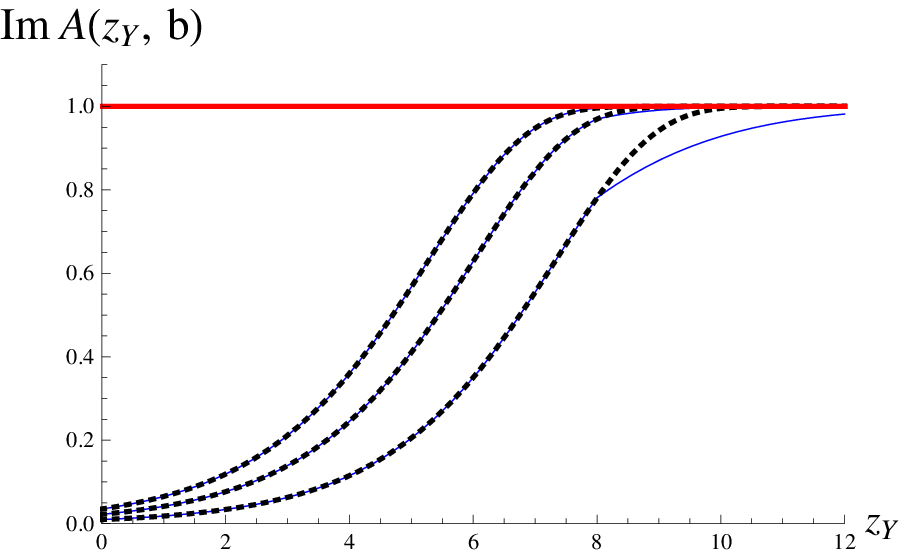,width=100mm}\\
\end{tabular}
\caption{The  scattering amplitude for gold-gold interaction versus $z \equiv z_Y$ at different values of $b$: the solid black line line is the result of this paper (see \eq{OM6}), the red dotted curve describes the Glauber-Gribov formula for one BFKL Pomeron exchange (see \eq{AA9}); and the red line shows the unitarity bound. The curves correspond $b = 0, 5\,fm$ and $10\,fm$ going from left to right,}
\label{ampaa}
}  
  
The main conclusions that we can make from \fig{ampaa} is the fact that the exact solution for nucleus-nucleus scattering shows the same  corrections as  Glauber-Gribov formula.  This happens because at $z_Y = 2 z_{min}$ the amplitude turns out to be very close to the unitarity bound.  However, it should be stress that actually the exact solution leads to slower approaching to the unitarity bound since it turns out (see \fig{omm}) that $\Omega\Lb z_Y, b \Rb$ of \eq{OMF}  is much smaller than $\Omega_P\Lb z _Y, b \Rb $ of \eq{OM4} that corresponds to the exchange of one BFKL Pomeron  at the entire kinematic region.

 \FIGURE[ht]{
\begin{tabular}{c c c }
\epsfig{file=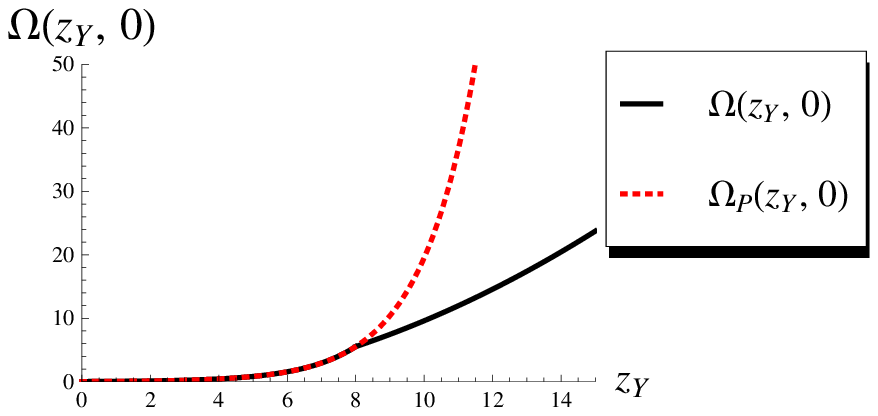,width=55mm}&\epsfig{file=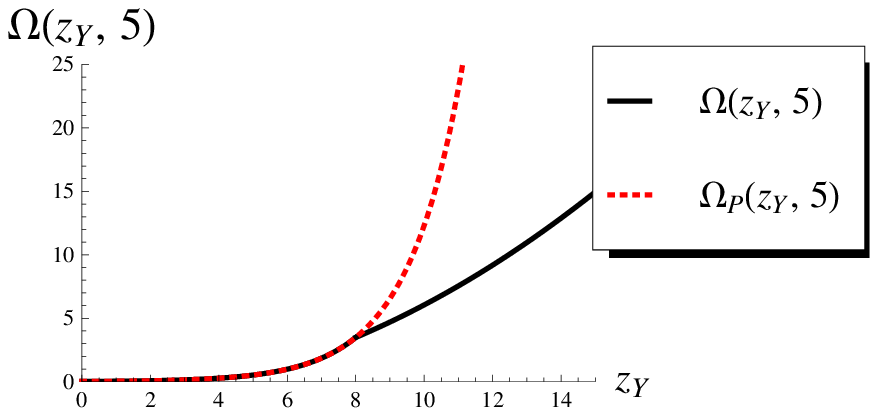,width=55mm} &\epsfig{file=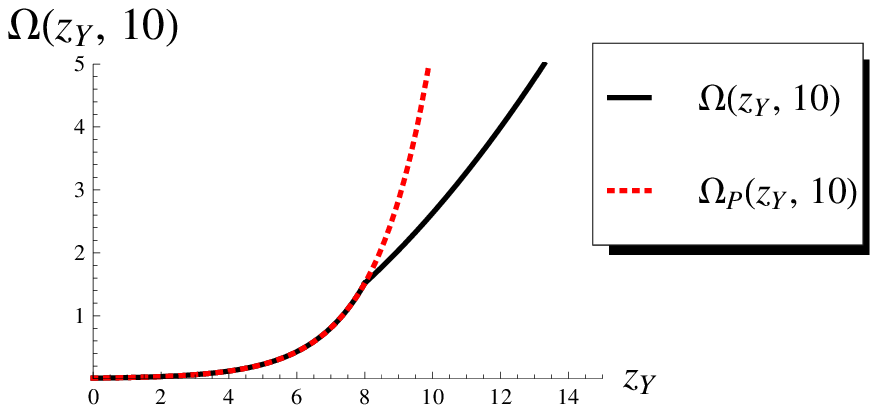,width=55mm}\\
 \end{tabular}
\caption{The  opacity $\Omega\Lb z_Y, b \Rb$ of \eq{OM6} and $\Omega_P\Lb z _Y\Rb$ of \eq{OM10}  for gold-gold interaction at different $b$ versus $z \equiv z_Y$: black line is the result of this paper (see \eq{OM4}), the red dotted curve describes 
 the contribution to $\Omega$ from exchange of one BFKL Pomeron. The values of $b$ are given in $fermi$.}
\label{omm}
}  
        
     \section{Conclusions}
The main result of this paper is the solution in the form of \eq{N} for amplitude $N\Lb Y,Y',l\Rb = N\Lb z, Y'\Rb$ with $\phi\Lb z,Y' \Rb$ given by \eq{SOLG} and \eq{LT1} and with the initial function $\phi^{(0)}\Lb z, Y' \Rb$ determined by \eq{ICAS}. 
We obtained this solution in two steps. First, we assume that \eq{NDAG}  ($ N\Lb l, b; Y; Y_0\Rb\,=\,N^\dag\Lb b; L - l, Y - Y_0\Rb$) which is correct for the linear equation, is valid for the solution of the non-linear one.
This conjecture allows us to reduce the system of two equations to one  functional differential equation.  Second, we solve this equation. The semi-classical approach was used to select out the infinite number of parasite solutions and the final solution was found analytically at large values of $Y'$ as well as numerically in the entire region of Y'.

The solution, that has been found, looks as  being different from the numerical solutions found in Refs.\cite{BOMO,BOBR}, but it shares with them several common features. In particular, this solution exists only and $z \leq z_{min} \approx 4$. In our solution the amplitude $N^\dagger\Lb Y,Y',l\Rb\,=\,N\Lb z_Y - z\Rb$.
We have no proof that this solution is unique but we believe that the physics problem has the only one solution.

Using this solution we found the nucleus-nucleus scattering amplitude as a function of energy for  `theoretical' nuclei. This 'theoretical' nucleus consists of the dipoles that are  made of heavy quarks and antiquarks. We can use the perturbative QCD treating this nucleus. In reconstruction of  the scattering amplitude we use   \eq{OMMODG} (see \fig{aaeq}) which is the form of $t$-channel unitarity constraints and follows from \eq{SE} for $S_E$ term in the action of \eq{BFKLFI}.

We showed that for $ z_Y\, \leq \,2 z_{min}$ the exact calculation give the same result as  the Glauber-Gribov formula which is widely used for nucleus-nucleus scattering. It turns out that $z_{min} \approx 4$ is so large that our estimates lead to the result close to the Glauber-Gribov formula in the entire kinematic region of energies. Only at large impact parameters the exact formula for nucleus-nucleus scattering gives visible deviations from the estimates based on Glauber-Gribov approach. We believe that this observation is important for all practical estimates based on Glauber-Gribov approach.

In spite of the fact that the solved problem is still far away from the real physics environment we hope that our solution gives a reasonable first approximation to approach the nucleus-nucleus scattering for the nuclei that exist in reality.

\section{Acknowledgements}
We thank all participants of the HEP seminar at UTFSM for useful discussions.
This research was supported by  the  Fondecyt (Chile) grants 1100648, 1095196 and  DGIP 11.11.05.

\end{document}